\documentclass[aps,prb,twocolumn,showpacs,superscriptaddress]{revtex4}
\usepackage[english]{babel}
\usepackage{amsmath,amssymb,bbm,graphicx,color,comment,txfonts}
\usepackage[bookmarks=true,colorlinks,citecolor=blue,urlcolor=blue]{hyperref}
\usepackage{dsfont}
\usepackage{float}

\newcommand{\Li}{LiHo$_x$Y$_{1-x}$F$_4$ }

\newcommand{\eq}[2]{\begin{eqnarray}\label{#1} #2 \end{eqnarray}}

\begin{document}

\date{\today}

\title{Quantum dynamics in strongly driven random dipolar magnets}
\author{M. Buchhold}
\affiliation{Department of Physics and Institute for Quantum Information and Matter, California Institute of Technology, Pasadena, CA 91125, USA}
\author{C.~S. Tang}
\affiliation{Department of Physics and Institute for Quantum Information and Matter, California Institute of Technology, Pasadena, CA 91125, USA}
\author{D.~M. Silevitch}
\affiliation{Department of Physics and Institute for Quantum Information and Matter, California Institute of Technology, Pasadena, CA 91125, USA}
\author{T.~F. Rosenbaum}
\affiliation{Department of Physics and Institute for Quantum Information and Matter, California Institute of Technology, Pasadena, CA 91125, USA}
\author{G. Refael}
\affiliation{Department of Physics and Institute for Quantum Information and Matter, California Institute of Technology, Pasadena, CA 91125, USA}

\begin{abstract}
The random dipolar magnet \Li enters a strongly frustrated regime for small Ho$^{3+}$ concentrations with $x<0.05$. In this regime, the magnetic moments of the Ho$^{3+}$ ions experience small quantum corrections to the common Ising approximation of \Li, which lead to a $\mathds{Z}_2$-symmetry breaking and small, degeneracy breaking energy shifts between different eigenstates.
Here we show that destructive interference between two almost degenerate excitation pathways burns spectral holes in the magnetic susceptibility of strongly driven magnetic moments in \Li. Such spectral holes in the susceptibility, microscopically described in terms of Fano resonances, can already occur in setups of only two or three frustrated moments, for which the driven level scheme has the paradigmatic $\Lambda$-shape. For larger clusters of magnetic moments, the corresponding level schemes separate into almost isolated many-body $\Lambda$-schemes, in the sense that either the transition matrix elements between them are negligibly small or the energy difference of the transitions is strongly off-resonant to the drive. This enables the observation of Fano resonances, caused by many-body quantum corrections to the common Ising approximation also in the thermodynamic limit. We discuss its dependence on the driving strength and frequency as well as the crucial role that is played by lattice dissipation.
\end{abstract}

%% I still have to check new associations in Fig1!

\maketitle
\section{Introduction} 
Magnetic dipoles with Ising symmetry randomly distributed on a lattice provide the opportunity to explore the effects of interactions\cite{Chak2004,Biltmo2007,Schechter2008}, disorder\cite{Brooke779}, frustration\cite{Gingras_2011}, random fields\cite{Silevitch2007,Schechter2008a,Tabei2008a}, entanglement\cite{Ghosh2003}, and quantum fluctuations\cite{Bitko1996}, with the ability to tune their interplay\cite{Reich1990}. 
 When driven out of equilibrium, new many-body states emerge, with characteristics that are the magnetic analogues to optically driven atomic systems, but involving numerous quantum degrees of freedom\cite{Ghosh2002a}. When decoupled from the thermal environment, the states are intrinsically non-linear with very small linewidths\cite{Schmidt2014}.   

Investigating the magnetic phases and dynamics in the disordered dipolar quantum magnet \Li has been the focus of this class of research activity for several decades\cite{Cooke_1975,MENNENGA198448,Gingras_2011,Hansen1975,Brooke779,Brooke2001,QuilliamSG,HeneliusNature}, yet explanations for several of its properties at low temperatures remain elusive \cite{Gingras_2011,Biltmo2007,Ghosh2003,jonsson2007}.  In \Li, the magnetic Ho$^{3+}$ cations mainly interact via dipole-dipole interactions. At large concentrations $x>0.3$ the magnetic dipoles form a quantum Ising magnet, with the possibility of applying an external transverse field to tune the quantum fluctuations\cite{Battison_1975,Beau1978,Bitko1996,Chak2004,Tabei2008}. Diluting the Ho$^{3+}$ concentration below $x<0.3$ gives rise to random frustration\cite{Silevitch2007,Biltmo2007,Tam2009,Pollack2014}, which leads to the formation of an Ising spin glass for $0.15\le x\le0.25$ at sufficiently low temperature $T\le 0.5$K and transverse field\cite{Brooke779,Ancona2008,Romitti2019}, while below concentrations of $x<0.15$ the nature of the low temperature state can be manifestly classical or quantum depending upon the strength of the thermal link to a heat bath\cite{Gingras_2011,Biltmo2008,Tam2009,Jon2008,Schmidt2014}. 

Recently, attention has been drawn to the dilute limit $x\le 0.05$ in which different experiments have observed aspects of an Ising spin glass\cite{Quilliam2007,QuilliamSG,HeneliusNature,Reich1990}, a quantum disordered, so-called ``antiglass'' with spin liquid characteristics \cite{Reich1987,Reich1990,Ghosh2002a,Ghosh2003} and isolated quantum degrees of freedom \cite{Silevitch2017}. The root of the irreconcilability of these observations seems to be found in the strength of the dissipation experienced by the magnet, i.e., by the coupling of the sample to the environment, as evidenced by recent experiments that tune a \Li sample from an Ising spin glass to an ``antiglass'' by reducing its thermal coupling to the environment\cite{Schmidt2014, Silevitch2017}.  

A key signature of the antiglass behavior is ``spectral hole burning'', i.e. the observation of a Fano resonance in the magnetic susceptibility $\chi(\omega)$ in a \Li sample, which is strongly driven by a time-dependent magnetic field\cite{Ghosh2002a,Silevitch2017}. Fano resonances are commonly a signature of quantum interference. In \Li the resonances are observable at arbitrary transverse fields and surprisingly small drive frequencies $\omega_d\approx 2\pi\times200$~Hz and probe frequency detunings $\omega_p-\omega_d\approx 2\pi\times2-5$~mHz. They occur if the \Li sample is well isolated from its environment and vanish if the coupling to the environment is increased. 

The magnetic moments in \Li form a complicated, disordered, and strongly interacting many-body problem, which is hard to address theoretically even in the simplified Ising approximation\cite{HeneliusNature,Andresen}. What is especially puzzling in the hole burning experiments \cite{Silevitch2017} is the presence of several, strongly separated energy scales, and the apparent sensitivity of hole burning to all of them. The dipole-dipole interaction between two neighboring moments is of the order of $\Delta V=500$~mK and falls off with a distance as $\sim 1/|\vec{r}|^{3}$. The \Li sample is held at a temperature of about $T=100$mK. The sample is driven by an oscillating magnetic field with Rabi frequency $\Omega_d\approx 2\mu$K, a drive frequency, which corresponds to $\omega_d\simeq 60$nK and its response is probed at a detuning $\omega_p-\omega_d\simeq0.6$pK. 

In this work, we start from a microscopic description of \Li, incorporating the full magnetic dipole-dipole interaction and the crystal field for the $J=8$ angular momentum state of each Ho$^{3+}$ ion. Using exact diagonalization, we show that the observation of Fano resonances can be explained on a qualitative level already for a single pair of Ho$^{3+}$-ions. The resonances appear as a consequence of interference between two quasi-degenerate excitation pathways, corresponding to a pair of quasi-degenerate quantum states, which can be coupled by applying an external, oscillating magnetic field.

In order to generalize this observation to more realistic samples with $n\ge 10$ magnetic degrees of freedom, we devise a toy model of effective spin-$\frac{1}{2}$ degrees of freedom, which captures the main ingredients for the observation of hole burning and reduces to the \Li Hamiltonian at low energies and for few magnetic moments. Exploring the dynamics of small samples shows that an external, oscillating magnetic field addresses only a small fraction of the many-body Hilbert space, for a given set of driving parameters. The predicted magnetic susceptibility $\chi(\omega)$ displays several spectral holes, which can be explained in terms of quasi-degenerate many-body excitation pathways and which match quantitatively very well with the experimental findings and energy scales. Within our simplified model, we can understand the origin and the importance of the different energy scales and, in addition, can find an explanation why the Fano resonance is only observed in the limit of very small coupling between the sample and the environment. 

Based on these findings, we propose an experimental scheme to manipulate the Fano signals by an external, acoustic drive of the lattice vibrations. The idea behind this approach is to engineer the dissipation rate of the magnetic moments by controlling their interactions with the phonon continuum. The latter is controlled by the number of phonons that are accessible for scattering at a given energy. Driving phonon modes explicitly generates a nonequilibrium phonon distribution, which is peaked at the drive frequency and increases the dissipation rate at matching energies. This reduces or even destroys the interference pattern of the Fano resonances. Observing this reverse or ``anti''-hole burning at the phonon drive frequencies would confirm our present explanation of hole burning and open a path to control the magnetic properties of \Li via both time-dependent magnetic fields and sound.

\section{Model}\label{SecII}
In this section we briefly review the microscopic model for the magnetic degrees of freedom in \Li compounds and illustrate that several aspects of the long time dynamics of the dilute material ($x\ll1$) are not captured by an effective Ising description. Instead, the strong dipole-dipole interaction between magnetic Ho$^{3+}$ atoms induces non-trivial entanglement in the magnetic degrees of freedom and lifts the expected $\mathds{Z}_2$ symmetry of an Ising magnet. We show that the deviation from a common, random Ising magnet becomes crucial at low temperatures or when the system is driven by an external field. For suitable driving frequencies, the latter resolves violations of the $\mathds{Z}_2$ symmetry and therefore the quantum nature of the magnet, which manifests itself via the absence of degenerate energy levels and the presence of non-vanishing moments $\langle \alpha| J^z|\beta\rangle\neq0$ for different eigenstates $|\alpha\rangle, |\beta\rangle$. 

\subsection{Microscopic Hamiltonian}
\Li is a magnetic material because of the magnetic Ho$^{3+}$ ions, in which the $4f^{10}$ electrons form an $^5I_8$ electronic ground state manifold\cite{Gingras_2011,Chak2004,Tabei2008}. In this manifold, each Ho$^{3+}$ ion $l$ is described by a $J=8$ total angular momentum degree of freedom $\vec{J}_l$. The coupling of each Ho$^{3+}$ ion to its non-magnetic neighbors via Coulomb interactions and ion-lattice coupling is described by a crystal field Hamiltonian $H_{\text{cf}}(\vec{J}_l)$ which aims to polarize $\vec{J}_l$ along the magnetic $c$-axis of the crystal. Exchange interactions between neighboring Ho$^{3+}$ are in general weak\cite{MENNENGA198448} and become negligible in the dilute limit $x\ll1$. The remaining interaction between two different Ho$^{3+}$ ions is the magnetic dipole-dipole interaction, which can, however, become rather strong due to the large total angular momentum $J=8$ carried by each Ho$^{3+}$. 

In addition to the mentioned terms, Ho-cations display a significant hyperfine interaction between the nuclear magnetic moment ($I=7/2$) and the electronic moments with a coupling constant $A_J=39$mK. At low temperatures $T<0.5$K, the hyperfine coupling renormalizes the effective low energy degrees of freedom and quantitatively modifies the phase diagram both for the ferromagnetic and for the spin glass transition\cite{Schechter2005,Schechter2008}. Qualitatively, however, the localized nuclear moments do not change the nature of the long-range coupled low energy degrees of freedom of the magnetic dipoles $\vec{J}$ away from a phase transition (as discussed in Sec. \ref{nuke}). As we show below, hole burning in \Li can be very well explained without considering hyperfine interactions. We will discuss potential modifications due to hyperfine interactions at the end of Sec.~\ref{SecIII}.

We consider no static external magnetic field $\vec{B}=0$, which yields the microscopic Hamiltonian\cite{Chak2004,Tabei2008}
\eq{Eq1}{
H=\sum_l H_{\text{cf}}(\vec{J}_l)+\frac{1}{2}\frac{\mu_0 g_L^2\mu_B^2}{4\pi}\sum_{l\neq m} L^{\alpha\beta}(\vec{r}_{lm}) J^\alpha_m J^\beta_l.
}
Here, $L^{\alpha\beta}(\vec{r})=\frac{\delta_{\alpha\beta}|\vec{r}|^2-3r^\alpha r^\beta}{|\vec{r}|^5}$ is the dipole-dipole matrix element between two Ho$^{3+}$ ions $l,m$, which is evaluated at their relative coordinate $\vec{r}=\vec{r}_{lm}=\vec{r}_l-\vec{r}_m$. The interaction strength depends on the Bohr magneton $\mu_B=\frac{2}{3}\frac{\text{K}}{\text{T}}$, the vacuum magnetic permeability $\mu_0=4\pi \ 10^{-7}\frac{\text{N}}{\text{A}^2}$ and the Land\'e $g$-factor $g_L=\frac{5}{4}$. 

\Li has a tetragonal structure with lattice constants $a=5.175$ \r{A} and $c=10.75$ \r{A} and four possible spots for Ho$^{3+}$, or Y$^{3+}$ ions per unit cell\cite{MENNENGA198448,Gingras_2011}. In terms of the unit cell coordinates $(a,a,c)$ their positions are at $(0,\frac{1}{2},\frac{3}{4}),\ (0,0,\frac{1}{2}),\ (\frac{1}{2},0,\frac{1}{4}),\ (\frac{1}{2},\frac{1}{2},0)$. This amounts to a minimal distance of $\Delta r_{\text{min}}\approx3.73$\r{A} between two Ho$^{3+}$ ions. The corresponding magnetic interaction energy is $A_{\text{dip}}=18.5$mK. 

For each Ho$^{3+}$ ion the crystal field Hamiltonian $H_{\text{cf}}(\vec{J}_l)$ features a two-fold degenerate ground state doublet and an excited singlet state separated by an energy $\approx 10.5$K from the ground states. The remaining $14$ eigenstates are separated more than $20$K from these three states. At low temperatures $T<10$K and zero magnetic field, thermal activation of excited states can be excluded and each magnetic moment is commonly projected onto the ground state manifold, i.e., treated as an Ising degree of freedom\cite{Gingras_2011}.

In the Ising approximation, each magnetic moment reduces to an Ising spin ${J_l^\alpha=\delta_{\alpha,z}C_{zz}\sigma_l^z}$\cite{ Tabei2008,Schechter2005,Biltmo2007}, whose orientation is described by the Pauli matrix $\sigma^z$. Under this transformation, the Hamiltonian \eqref{Eq1} reduces to $H\rightarrow H_{\text{Ising}}$ with
\eq{Eq2}{
H_{\text{Ising}}=\frac{A_{\text{dip}}C_{zz}^2}{2}\sum_{l\neq m} \sigma^z_l\sigma^z_m L^{zz}({R_{lm})}.
}
Here $A_{\text{dip}}$ is the coupling introduced above, $C_{zz} \approx 5.5$ is the effective magnetic moment in the $z$-direction
and $R_{lm}$ is the dimensionless distance between spin $l$ and $m$ in units of $\Delta r_{\text{min}}$.

In the dense limit $x\ge 0.25$, the Hamiltonian \eqref{Eq2} describes an Ising dipolar ferromagnet with ordering temperature $T_c=1.53$K at $x=1$\cite{Chak2004,Cooke_1975,Beau1978,Brooke779}. For stronger dilution, $x<0.25$, the random positions of the Ho$^{3+}$ ions induce frustration between different magnetic moments and the system enters a dipolar Ising spin glass phase at sufficiently low temperatures $T<T_c\approx x T_c^{\text{mf}}=x \cdot 1.5$K\cite{Ancona2008,Quilliam2007,QuilliamSG,Tam2009,HeneliusNature}.  

There was a debate in the literature, whether \Li enters a glass state at very low concentrations $x\le 0.05$ and temperatures $T\le 100$mK. While some experiments showed strong evidence of a dipolar spin glass in this regime\cite{Quilliam2007,QuilliamSG}, which is supplemented by classical Monte Carlo simulations\cite{HeneliusNature} of the Ising Hamiltonian \eqref{Eq2}, another set of experiments reported evidence for an anti-glass state in which low energy quantum fluctuations prevent the spin glass freezing\cite{Reich1987,Reich1990,Ghosh2002a,Ghosh2003}. A most recent experimental study showed that both, glass and anti-glass behavior can be realized in the same setup by changing the systems interaction with the environment from strong (glass state) to weak (anti-glass state) coupling\cite{Silevitch2017}. 

Colloquially speaking, the Hamiltonian \eqref{Eq1} can be well approximated by the Ising Hamiltonian \eqref{Eq2} for $x\le 0.05$, if the dissipation rates, corresponding to dephasing and incoherent flips of the magnetic moments, are larger than the energy level splittings between quasi-degenerate states. In this case, dissipation dominates over coherent dynamics and the dynamics looks effectively classical, i.e. Ising-like. This statement will be made more quantitative below by showing that the hole burning, associated to the anti-glass dynamics, can be explained on the basis of  the Hamiltonian \eqref{Eq1} {\bf{without}} performing the Ising approximation. 

\subsection{Dimer and Trimer level schemes}
In order to point out the importance of quantum effects in the \Li Hamiltonian in \eqref{Eq1} for $x\le 0.05$, we compare the low energy physics of $H$ with $H_{\text{Ising}}$ for small spin 'clusters' of $n=2,3$ magnetic moments and refer to $n=2, 3$ as a dimer, trimer setup. We choose $n$ positions for the Ho$^{3+}$ atoms and diagonalize the $17^n\times 17^n$ Hamiltonian $H$ with exact material parameters\cite{Ronnow2007}. The precise form of the crystal field Hamiltonian for \Li is discussed in the Appendix \ref{CFH}.

\begin{figure}\center\includegraphics[width=\linewidth]{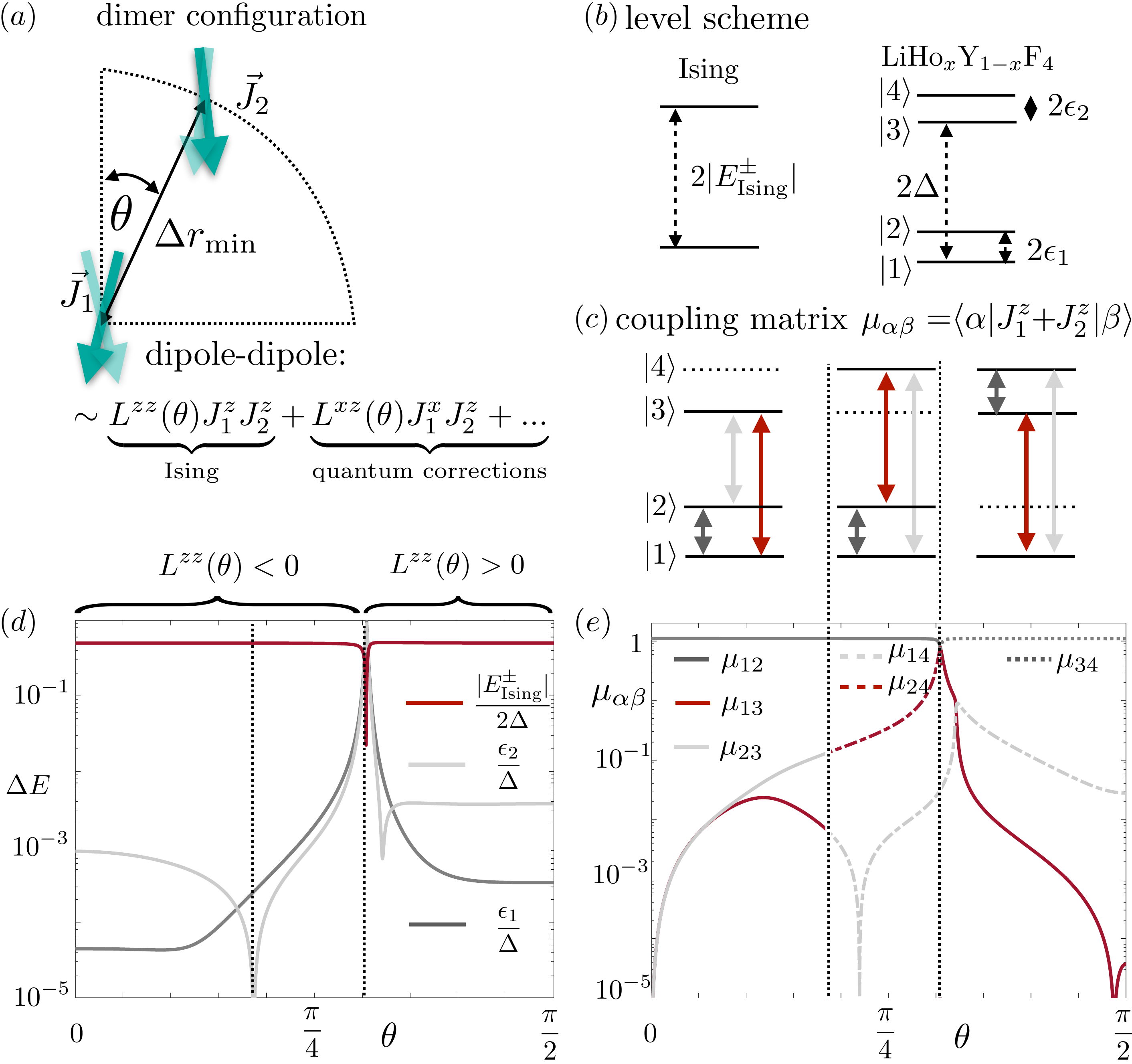}\caption{A magnetic dimer is the most simple unit from which hole burning in \Li can be understood. (a) It is formed by two $J=8$ magnetic moments, which interact with the material's crystal field and experience a mutual dipole-dipole force. The crystal field forces each magnetic moment to align along the $z$-axis and features an Ising-type ground state manifold associated with spin up and down. This singles out the Ising contribution $\sim J^z_1J^z_2$ as the dominant dipole-dipole interaction at low temperatures. Quantum corrections, led by the terms $\sim J^x_1J^z_2, J^z_1J^x_2$, are strongly suppressed by the crystal field but crucial for the understanding of hole burning in driven \Li samples. (b) The terms $~J^z_1J^x_2+...$ are not compatible with the Ising-symmetry. They lift the Ising degeneracy and introduce small level splittings $\epsilon_{1,2}$ between two quasi-degenerate eigenstates in the \Li dimer. (c) Breaking the Ising symmetry also introduces small but non-zero transition matrix elements $\mu_{\alpha\beta}=\langle\alpha|(J^z_1+J^z_2)|\beta\rangle$ between different dimer eigenstates $|\alpha\rangle, |\beta\rangle$. 
(d+e) Quantitative analysis of the level spacings (d) and transition matrix elements (e) from exact diagonalization of a \Li dimer described by $H^{(2)}$ in Eq.~\eqref{BEq1} with relative orientation, $\vec{r}_{12}=\Delta r_{\text{min}}(\sin\theta,0,\cos\theta)$ (we set $L^{\alpha\beta}(\theta)\equiv L^{\alpha\beta}(\theta)(\vec{r}_{12})$. The energies are compared to an equivalent Ising dimer, described by Eq.~\eqref{BEq2}. The colors in (e) match with the illustration in (c). At $\theta=\arccos\frac{1}{\sqrt{3}},  \frac{3\pi}{16}$, the states $|3\rangle$ and $|2\rangle$, $|4\rangle$ are degenerate. For some $\theta$, there is one "dark" state (dashed line) corresponding to an Ising singlet, which does not couple to the other states via $J^z$. The quasi-degenerate partner of the dark state, however, weakly couples to both states of the remaining quasi-degenerate pair. We refer to the particular form of $\mu_{\alpha\beta}$ in (c,e), i.e. the coupling of a quasi-degenerate pair of states to an energetically well separated state, as "$\Lambda$-scheme". It is the basic building block for hole burning in driven \Li.}\label{Fig1}\end{figure}%%%

First, we consider a dimer setup of two Ho$^{3+}$ ions with  $J=8$, which both experience the crystal field and mutual dipole-dipole interactions. The Hamiltonian of the two ions $l=1,2$, which are separated by a vector $\vec{R}_{12}$ (in units of $\Delta r_{\text{min}}$), is
\eq{BEq1}{
H^{(2)}=H_{\text{cf}}(\vec{J}_1)+H_{\text{cf}}(\vec{J}_2)+ A_{\text{dip}} \sum_{\alpha,\beta=x,y,z}L^{\alpha\beta}(\vec{R}_{12}) J_1^\alpha J_2^\beta.
}
The corresponding Ising Hamiltonian is obtained by projecting onto the ground state doublets of $H_{\text{cf}}(\vec{J}_{1,2})$ and is
\eq{BEq2}{
H_{\text{Ising}}^{(2)}=A_{\text{dip}}C_{zz}^2L^{zz}(\vec{R}_{12})\sigma^z_1\sigma^z_2.
}
It has eigenenergies $\pm E_{\text{Ising}}=\pm |A_{\text{dip}}C_{zz}^2L^{zz}(\vec{R}_{12})|$, each of which are two-fold degenerate.

In general, the dipole-dipole interaction in Eq.~\eqref{BEq1} does not feature a compatible $\mathds{Z}_2$-symmetry and thus breaks the ground state degeneracy of the crystal field Hamiltonian. This introduces splitting energies $\epsilon_{1,2}$ as illustrated in Fig.~\ref{Fig1}(b). We introduce the projector $P^{(n)}$, which projects onto the $2^n$ states of lowest energy of $H^{(n)}$. For each dimer eigenbasis one finds
\eq{BEq3}{
P^{(2)}H^{(2)}P^{(2)}&=&2\ \text{diag}(\Delta+\epsilon_2,\Delta,\epsilon_1,0),\\
H_{\text{Ising}}^{(2)}&=&2\ \text{diag}(E_\text{Ising},E_\text{Ising},0,0).
}

Away from the special point $L^{zz}(\vec{R}_0)=0$, where the 'classical Ising' interaction vanishes, the modifications of the eigenvalues of $P^{(2)}H^{(2)}P^{(2)}$ compared to $H^{(2)}_{\text{Ising}}$ seem rather small, i.e., $\frac{\epsilon_{1,2}}{\Delta}, \left|\frac{\Delta-E_{\text{Ising}}}{E_{\text{Ising}}}\right|\sim 10^{-3}- 10^{-4}$ for $|\vec{R}|=1$, see Fig.~\ref{Fig1}(d). For $|\vec{R}|>1$, one finds a very accurate scaling estimate
\eq{BEq4}{
\left. \frac{\epsilon_{1,2}}{\Delta}\right|_{|\vec{R}|>1}\approx \left. \frac{\epsilon_{1,2}}{\Delta}\right|_{|\vec{R}|=1}\frac{1}{|\vec{R}|^{3}}. 
}
This anticipates that corrections of $P^{(2)}H^{(2)}P^{(2)}$ compared to $H_{\text{Ising}}^{(2)}$ can be understood in terms of second order perturbation theory. The eigenvalues of the dipole matrix $J^\alpha_1J^\beta_2$ can, however, become very large. Using second order Brillouin-Wigner perturbation theory \cite{} in the eigenbasis of $H_{\text{cf}}(\vec{J}_1)+H_{\text{cf}}(\vec{J}_2)$ we find that convergence towards $P^{(2)}H^{(2)}P^{(2)}$ requires to include more than $N=100$ of the $17^2=289$ eigenstates. This makes it difficult to express the eigenstates of  $P^{(2)}H^{(2)}P^{(2)}$ in the Ising basis analytically.

A second modification caused by using $H^{(2)}$ instead of $H^{(2)}_{\text{Ising}}$ is that the total $z$-axis magnetization $J_{\text{tot}}^z=J_{1}^z+J_2^z$ is no longer diagonal in the basis of energy eigenstates. For the Ising Hamiltonian, $J^z_l\propto\sigma^z_l$ and $[\sigma^z_l,H_{\text{Ising}}]=0$ and $J_{\text{tot}}^z, H_{\text{Ising}}$ can be diagonal in the same basis. 
In contrast, all diagonal matrix elements of $J^z_{\text{tot}}$ vanish in the eigenbasis of $H$. We define the matrix elements of the total magnetic moment in the $z$-direction between eigenstates $|\alpha\rangle, |\beta\rangle$,
\eq{BEq5}{
\mu_{\alpha\beta}=\langle \alpha | J^z_{\text{tot}}|\beta\rangle=\sum_l \langle \alpha| J^z_l |\beta\rangle.
} 
For Ising eigenstates, we find $\mu_{\alpha\beta}\sim\delta_{\alpha\beta}$, while for \Li clusters we find $\mu_{\alpha\beta}\sim(1-\delta_{\alpha\beta})$. The absolute values $|\mu_{\alpha\beta}|$ for the dimer setup are shown in Fig.~\ref{Fig1}(e). 

For non-commuting $[J^z_{\text{tot}},H]\neq0$, an oscillating magnetic drive field $h(t)=h_d\cos(\omega_dt)$ in the $z$-direction, which is described by the Hamiltonian
\eq{BEq6}{
\delta H(t)=h_d\cos(\omega_dt)\mu_B\mu_0 J^z_{\text{tot}},
}
induces transitions between different energy eigenstates $|\alpha\rangle \leftrightarrow|\beta\rangle$. The transition rates are proportional to $|\mu_{\alpha\beta}|$, see ~\ref{Fig1}(e), and the corresponding level schemes for the dimer setup are illustrated in Fig.~\ref{Fig1}(c). The transition matrix $\mu_{\alpha\beta}$, which couples a quasi-degenerate pair of states to another, energetically well separated state, has the shape of a (inverse) $\Lambda$ and we refer to it as $\Lambda$-scheme. As we will discuss, it features a similar dynamics as driven three-level systems in quantum optics, where $\Lambda$-schemes of this shape are common.  The $\Lambda$-scheme is the basic building block for the understanding of hole burning in \Li and we will analyze it in detail in Sec.~\ref{SecII}.

Adding more magnetic moments to the cluster either enhances or suppresses corrections to the Ising approximation and may lead to more involved coupling matrices $\mu_{\alpha\beta}$. We demonstrate this in the following by analyzing magnetic trimer configurations with at least one frustrated moment. The corresponding \Li- and Ising- Hamiltonians are
\eq{BEq7}{
H^{(3)}&=&\sum_{l=1}^3 H_{\text{cf}}(\vec{J}_l)+A_{\text{dip}}\sum_{\alpha,\beta=x,y,z}\sum_{l=1, m>l}^3L^{\alpha\beta}(\vec{R}_{lm})J^\alpha_lJ^\beta_m,\ \ \ \ \ \\
H^{(3)}_{\text{Ising}}&=&A_{\text{dip}}C_z^2\sum_{l=1,m>l}^3L^{zz}(\vec{R}_{lm})\sigma^z_l\sigma^z_m.
}

\begin{figure}\center\includegraphics[width=\linewidth]{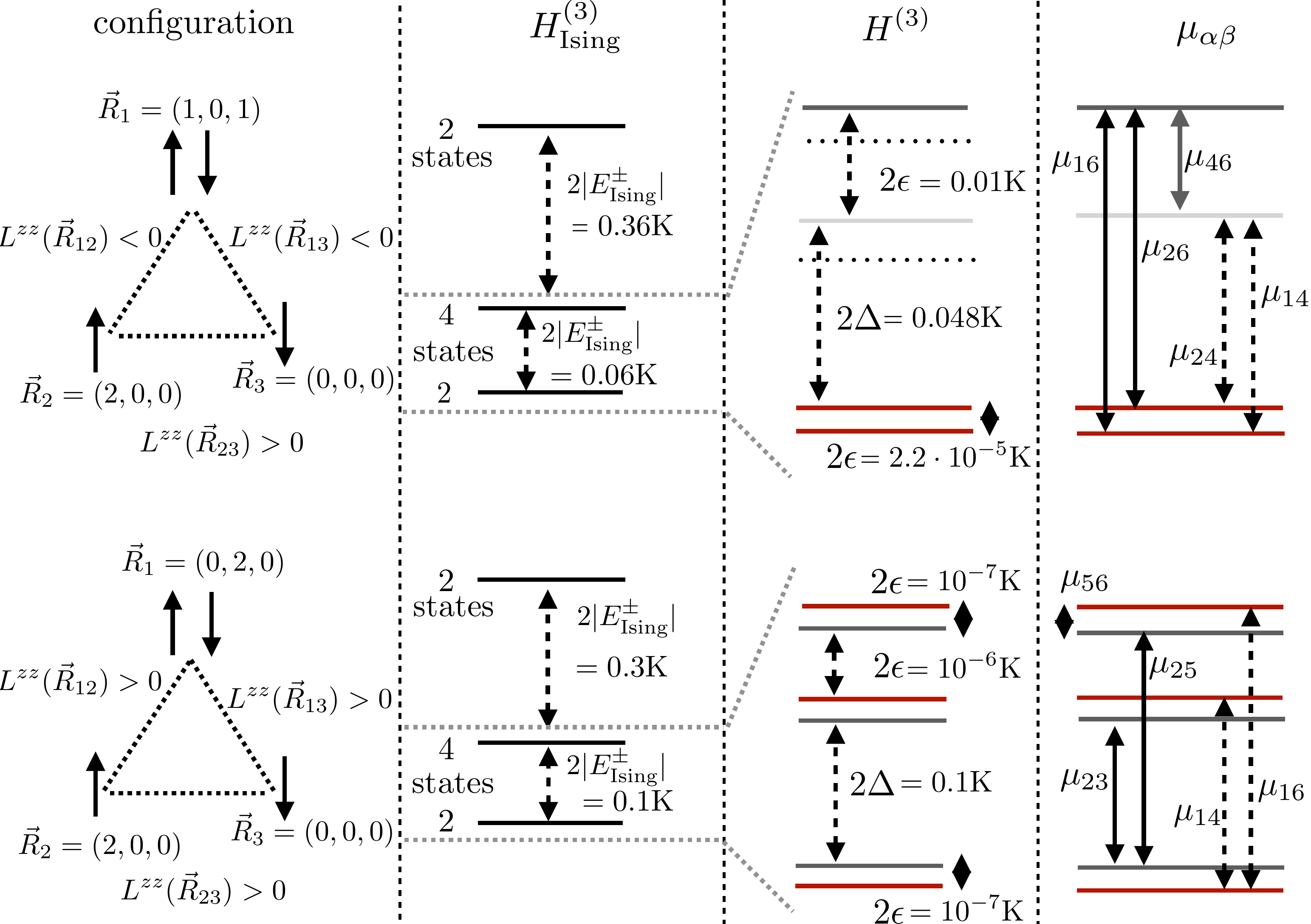}\caption{Level scheme and transition matrix elements for the low energy eigenstates of two generic \Li trimer configurations. Among the 3 Ising couplings in the top (bottom) row,  2 (0) are ferromagnetic and 1 (3) are antiferromagnetic. As for the \Li dimer, the trimer energy scheme (3rd column) shows several quasi-degenerate level splittings compared to the degenerate Ising scheme (2nd column). The ratio of a quasi-degeneracies over its corresponding Ising energy ranges from $|\epsilon/E^{\pm}_{\text{Ising}}|\approx 0.2-10^{-6}$ and covers a wider range than a dimer with comparable distances between the magnetic moments. The transition matrix $\mu_{\alpha\beta}=\langle\alpha|\sum_{l=1}^3 J^z_l|\beta\rangle$ (right column) for each trimer configurations can be decomposed into two distinct $\Lambda$-schemes corresponding to solid and dashed arrows, each of which corresponds to a characteristic energy difference $\Delta$ and quasi-degeneracy $\epsilon$. The quasi-degeneracies in the spectrum and the composition of the transition matrix $\mu_{\alpha\beta}$ from $\Lambda$-schemes are genuine features of magnetic clusters in \Li, originating from the weak breaking of an effective Ising symmetry in the crystal field's ground state manifold. 
}\label{Fig2}\end{figure}

The level scheme for two specific trimer configurations is shown in Fig.~\ref{Fig2}. The deviations of the exact level scheme from the one predicted by the Ising approximation, i.e., the degeneracy breaking energies $\epsilon$, range from very small values $\epsilon\sim 10^{-7}$K to relatively large ones $\epsilon\sim 0.01$K, depending on the spatial configuration of the magnetic moments. In any case, one finds a complex matrix structure of $\mu_{\alpha\beta}$, shown in Fig.~\ref{Fig2} (right column), as compared to the diagonal structure predicted by the Ising approximation. 

When the system is driven with an time-dependent magnetic field, i.e., when adding $\delta H(t)$ in Eq.~\eqref{BEq6}, the transition matrix $\mu_{\alpha\beta}$ corresponding to $H^{(n)}$ enables coherent, magnetization changing transitions between the states $|\alpha\rangle \leftrightarrow |\beta\rangle$ with Rabi frequency $\Omega_{\alpha\beta}=h_d\mu_B\mu_0\mu_{\alpha\beta}$. This is in contrast to the classical Ising Hamiltonian, which remains diagonal in the presence of a magnetic field in the $z$-direction, i.e., does not induce coherent transitions between different Ising eigenstates. The consequences of the particular structure of $\mu_{\alpha\beta}$ in \Li for the response to external driving, in particular, how it leads to hole burning, will be discussed in the following section. 

%%%

\section{Hole burning in driven $Li Ho_x Y_{1-x} F_4$ dimers and trimers}\label{SecII}
Including the quantum corrections to the Ising Hamiltonian, the phenomenon of spectral hole burning, i.e., the emergence of a Fano resonance in the magnetic susceptibility $\chi(\omega)$, can be explained theoretically even in the most simple dimer and trimer level schemes for a \Li Hamiltonian discussed above. The origin of the Fano resonance in a dimer or trimer is quantum interference between two almost degenerate excitation pathways. We will discuss this phenomenon on the basis of the instructive dimer scheme and highlight the role played by dissipation for the resonance in this section, before we discuss its generalization to the case of many moments in the following section.  

\subsection{Magnetic dissipation rates}
In order to study the dynamics of a dimer under external driving, we need some estimate on the environmental induced dissipation, i.e., the dissipative transition rates $\gamma_{\alpha\rightarrow\beta}$ between two quantum states $|\alpha\rangle, |\beta\rangle$. The major source of dissipation for the magnetic moments in \Li is the coupling of the Ho$^{3+}$ ions to lattice vibrations, i.e., phonons. In order to estimate the associated dissipation rates, we consider acoustic, Debye type low energy phonon modes, which are described by a wavevector $\vec{k}$, dispersion $\omega_{\vec{k}}=c|\vec{k}|$, Debye frequency $\omega_D$ and Debye temperature $\Theta_D$. 

Each phonon mode has a linewidth $\Gamma$, which describes the rate at which the mode exchanges energy with other phonons and the environment. Due to weak phonon-phonon interactions, the linewidth is dominated by the coupling of the sample to the environment. In recent experiments\cite{Schmidt2014,Silevitch2017} it has been pointed out that having a weak sample-environment coupling is crucial for the observation of hole burning and antiglass dynamics in driven \Li. Here, we consider the lattice-environment coupling in terms of an effective phonon linewidth. Such a linewidth will set the lower bound for the magnetic dissipation rates, and enable or disable the emergence of a Fano resonance. This yields a phenomenological explanation for the presence or absence of hole burning in several \Li experiments.

Acoustic phonons in the Debye model are described by the Hamiltonian
\eq{BEq8}{
H_D=\sum_{\vec{k}, c|\vec{k}|<\omega_D}c|\vec{k}| b^\dagger_{\vec{k}}b^{\phantom{\dagger}}_{\vec{k}},
}
with bosonic ladder operators $b^\dagger_{\vec{k}}, b^{\phantom{\dagger}}_{\vec{k}}$ at momentum $\vec{k}$. The linear coupling between the phonons and the magnetic states is typically of the form
\eq{BEq9a}{
H_{\text{mag-ph}}=\sum_{\alpha,\beta,\vec{k}}g_{\alpha\beta}(\vec{k})\left(b^\dagger_{\vec{k}}+b^{\phantom{\dagger}}_{-\vec{k}}\right)\left(|\alpha\rangle\langle\beta|+|\beta\rangle\langle\alpha|\right),
}
with coupling matrix elements $g_{\alpha\beta}(\vec{k})$ between different eigenstates $|\alpha\rangle, |\beta\rangle$ of the magnetic Hamiltonian \eqref{Eq1}.

The transition rate $\gamma_{\alpha\rightarrow\beta}$ between two magnetic states $|\alpha\rangle\rightarrow |\beta\rangle$ with energy difference $E_{\alpha\beta}\equiv E_\alpha-E_\beta$ can be estimated by Fermi's golden rule (see Appendix~\ref{LatDis}). It yields an energy-dependent decay rate $\gamma_{\alpha\beta}=\gamma(E_{\alpha\beta})=\gamma(E)$
\eq{BEq8a}{
\gamma(E)&=&-i 
%\sum_{\vec{k}, \{n_{\vec{k}}, n_{\vec{k}}'\}} \frac{|\langle \alpha, n_{\vec{k}}| H_{\text{sp-ph}}|\beta, n_{\vec{k}}'\rangle|^2}{\Delta_{\alpha\beta}+\omega_{\vec{k}}(n_{\vec{k}}-n_{\vec{k}}')}=
\sum_{{\vec{k}}}|g(E,\vec{k})|^2\left(\frac{n_{\vec{k}}}{E+i0^++c|\vec{k}|}+\frac{n_{\vec{k}}+1}{E+i0^+-c|\vec{k}|}\right)\nonumber\\
&=&|g(|E|)|^2 \left(n_B(|E|)+\delta_{\text{sign}(E),1}\right)\rho_{\text{ph}}(|E|).
}
Here $n_B(|E|)$ is the Bose-Einstein distribution at temperature $T$ and energy $|E|$, $\rho_{\text{ph}}(E)$ is the phonon density of states and we used the shortcut $g(|E|)\equiv g_{\alpha\beta}(c|\vec{k}|=E_{\alpha\beta})$. 

For energy differences $E\ll k_BT$, the Bose function shows the typical $\frac{1}{E}$-divergence $n_{B}(|E|)\approx \frac{k_B T}{|E|}\gg1$. For acoustic phonons at low energies, $g(|E|)=g_0 \sqrt{|E|}$, which yields $\gamma(E)\approx g_0^2 k_B T \rho_{\text{ph}}(|E|)=\gamma_D\frac{T\rho_{\text{ph}}(|E|)}{\Theta_D\rho_{\text{ph}}(\omega_D)}$, where $\gamma_D$ is the decay rate at the Debye frequency.  For a linear dispersion with linewidth $\Gamma$ the density of states in $d=3$ dimensions is well approximated by $\rho_{\text{ph}}(|E|)=\rho_0 E^2$ for $E\gg\sqrt{\Gamma\omega_D}$ and a constant $\rho_{\text{ph}}(|E|)=\rho_0 \Gamma\omega_D$ for $E\ll\Gamma$. This yields the dissipation rate
\eq{BEq9}{
\gamma(|E|)\approx\gamma_D\frac{T}{\Theta_D\omega_D^2} \times\left\{\begin{array}{cc} E^2& \text{for }\sqrt{\Gamma\omega_D}<E\\ \frac{\Gamma\omega_D}{2\pi}& \text{for } \sqrt{\Gamma\omega_D}> E\end{array}\right. . 
}

One thus observes that for very small $\Gamma, E$, the dissipation rates, which push the system back towards its equilibrium state can become very small, leading to an out-of-equilibrium state under driving. For \Li the parameters in Eq.~\eqref{BEq9} are hard to quantify due to the lack of knowledge on the interaction between phonons and the magnetic moments in the material. In order to obtain a qualitative estimate, one might consider the dissipation rates of spin vacancies in diamond, where the phonon-induced dissipation has been determined very precisely to be $\frac{\gamma_D}{\Theta_D\omega^2_D}=10^{-15}-10^{-14}\frac{1}{\text{Hz K}}$ \cite{Alp2012,Pingault}\footnote{Determined from the spin coherence times at temperatures $T\approx 8K$ and phonon energies $E=40$GHz.}. For the significantly small level spacings of $E\sim 1\text{kHz}$, which are addressed by the driving field, the dissipation is dominated by the phonon linewidth, yielding $\gamma\approx\frac{\Gamma T}{100K}$ with the Debye temperature of LiHoF$_4$ being $\Theta_D=600K$. The linewidth $\Gamma$ sets a lower bound to the dissipation rate, indicating that for strong coupling to the environment, the system is hardly pushed away from its equilibrium. 

\begin{figure}
\center
\includegraphics[width=1\linewidth]{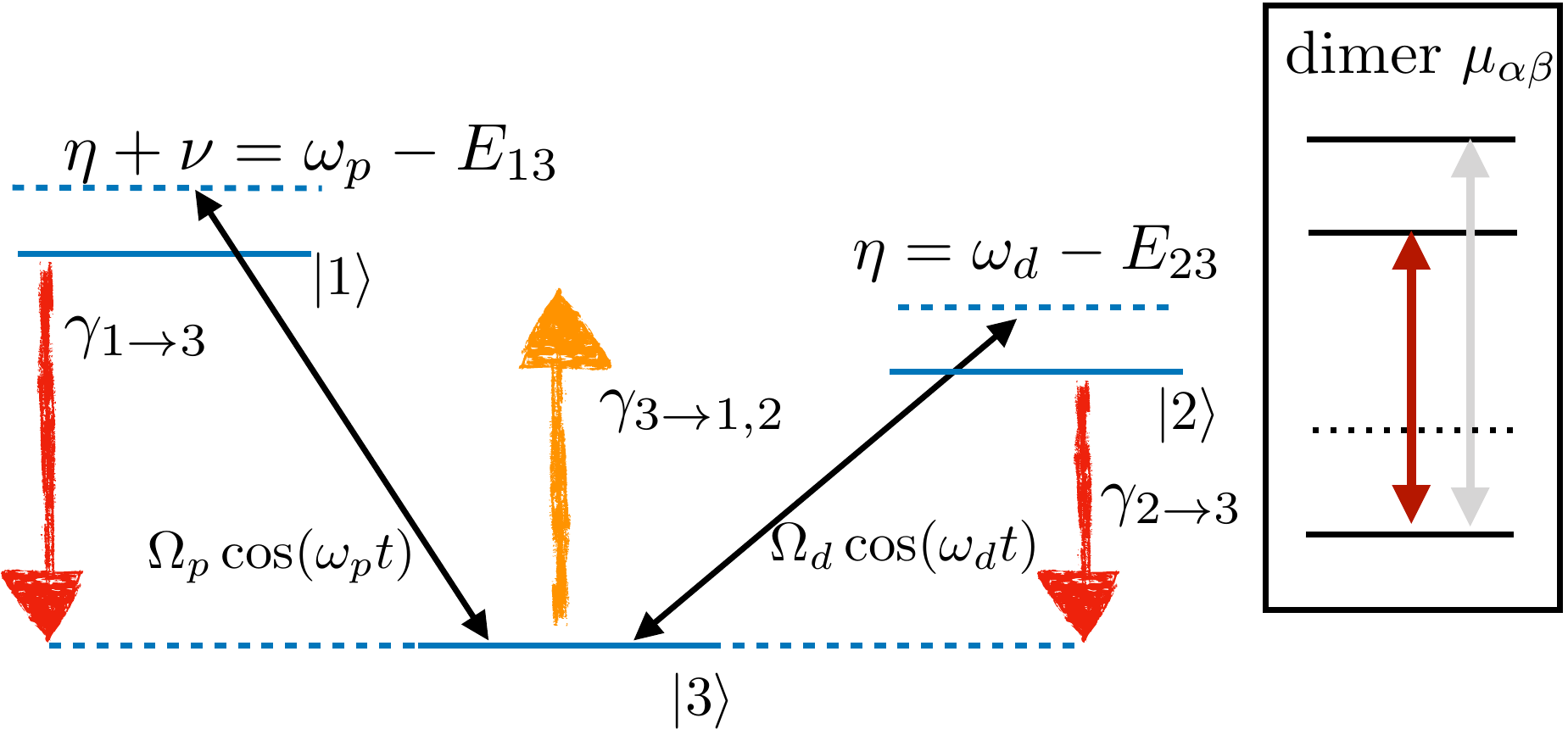}
\caption{Illustration of an idealized, driven $\Lambda$-scheme, which is realized in an antiferromagnetic Ho$^{3+}$ dimer or trimer subject to time-dependent magnetic drive and probe fields in the $z$-direction. The inset shows a corresponding dimer transition matrix $\mu_{\alpha\beta}$ extracted from Fig.~\ref{Fig1} (c). The drive, probe fields here oscillate with frequency $\omega_d, \omega_p$ and the strength of the couplings is described by the Rabi frequency $\Omega_d, \Omega_p$, which is proportional to the corresponding coupling matrix elements and the strength of the magnetic drive, probe fields $h_p, h_d$, i.e., $\Omega_p\propto h_p\mu_{23}, \Omega_d\propto h_d\mu_{13}$.  The drive scheme also includes dissipative transitions with rates $\gamma_{\alpha\rightarrow\beta}$, corresponding to Stokes ($\uparrow$) and anti-Stokes ($\downarrow$) transitions, which stem from the coupling of the magnetic moments to a low temperature phonon continuum. Adjusting the detuning $\eta, \nu+\eta$ of the drive and probe field from the energy differences $E_{13}, E_{23}$ in the $\Lambda$-scheme enables a Fano resonance, i.e., hole burning,  in the linear susceptibility $\chi$. Note: This illustration represents an idealization. In reality, both the drive and probe fields contribute to $\mu_{13}$ and $\mu_{23}$ at the same time. In linear response to the probe field $h_p$, however, the measured signal is very well approximated by $\Lambda$-schemes linear in $h_d$. In addition to this figure, this is covered by a drive scheme with $(\omega_d,h_d)\leftrightarrow(\omega_p,h_p)$. 
}\label{Fig3}\end{figure}%%%

\subsection{Magnetic susceptibility and Fano signal for spin dimers}\label{SecIIB}
The $\Lambda$-schemes found in magnetic dimers and trimers in Figs.~\ref{Fig1}(c) and \ref{Fig2} are common candidates for the observation of interference between different excitation pathways and Fano resonances\cite{Fleischi2005,Limonov2017}. In this section, we discuss the mechanism of destructive interference, which leads to a Fano resonance in the magnetic susceptibility, for an idealized $\Lambda$-scheme. The $\Lambda$-scheme is illustrated in Fig.~\ref{Fig3}. It consists of three quantum states $|1,2,3\rangle$, which are driven by two external fields. The $|2\rangle\leftrightarrow |3\rangle$ transition is driven by a time-dependent driving field and the $|1\rangle\leftrightarrow |3\rangle$ transition is driven by a time-dependent probe field. The measured time-dependent magnetic susceptibility will be proportional to the coherences $|3\rangle\langle 2|, |3\rangle\langle1|$. Their dynamics does not depend on whether the $\Lambda$-scheme is regular or inverted and without loss of generality, we discuss an inverted scheme. The generalization to the situation of many magnetic moments follows in Sec.~\ref{SecIII}.

The ideal $\Lambda$-scheme consists of three levels $|l\rangle, l=1,2,3$ corresponding e.g., to three different dimer eigenstates, as shown in Fig.~\ref{Fig3}(c). An oscillating external magnetic field with Rabi frequency $\Omega_d$ and drive frequency $\omega_d$ drives the $|2\rangle\leftrightarrow|3\rangle$ transition with a detuning $\eta=\omega_d-E_{23}$ from resonance. At the same time, an oscillating probe field with Rabi frequency $\Omega_p$ and drive frequency $\omega_p$ probes the $|1\rangle\leftrightarrow|3\rangle$ transition with detuning $\nu+\eta=\omega_p-E_{13}$ from resonance. In addition, incoherent transitions are induced by the coupling of the states to a phonon continuum. The corresponding rates for the (anti-) Stokes $\gamma_{3\rightarrow1,2}$ ($\gamma_{1,2\rightarrow3}$) processes can be estimated by Eq.~\eqref{BEq9}.

The time-dependent Hamiltonian for this $\Lambda$-scheme is 
\eq{BEq10}{
H_\Lambda(t)&=&E_{13} |1\rangle\langle 1|+E_{23} |2\rangle\langle 2|+\Omega_d\cos(\omega_d t)(|2\rangle\langle 3|+\text{h.c.})\nonumber \\
&&+\Omega_p\cos(\omega_p t)(|1\rangle\langle 3|+\text{h.c.})
}
with $E_{13}, E_{23}>0$. Assuming small Rabi frequencies, $\Omega_{d,p}\le \omega_{d,p}$, one can perform a rotating wave approximation (RWA), i.e., transform the $H_V(t)$ into a frame rotating with the drive and pump fields and discard all counterrotating terms $\sim 2\omega_{p,d}$. The corresponding unitary transformation is $U(t)=\exp\left[it\left(\omega_p |1\rangle\langle1|+\omega_d |2\rangle\langle2| \right)\right]$ and the transformed Hamiltonian 
\eq{BEq11}{
\tilde{H}_{\Lambda}&=&U^\dagger(t)H_\Lambda(t)U(t)-iU^\dagger(t)\partial_tU(t)\nonumber\\
&\overset{\text{RWA}}{=}&\frac{\Omega_d}{2}(|3\rangle\langle2|+|2\rangle\langle3|)+\frac{\Omega_p}{2}(|3\rangle\langle1|+|1\rangle\langle3|)\nonumber\\
&&+\nu|2\rangle\langle2|+(\nu+\eta)|3\rangle\langle3|.
}
In the last step, we added a constant energy shift $\tilde{H}_\Lambda\rightarrow\tilde{H}_\Lambda+(\nu+\eta)$, which does not change the dynamics.

In order to account for the dissipation, we use a density matrix description of the magnetic system. The density matrix
\eq{BEq12}{
\hat{\rho}=\sum_{\alpha,\beta=1}^3\rho_{\alpha\beta}|\alpha\rangle\langle\beta|
}
is hermitian $\rho_{\alpha\beta}=\rho_{\beta\alpha}^*$ and has unit trace $\sum_{\alpha}\rho_{\alpha\alpha}=1$. Its time evolution is described by a quantum master equation in Lindblad form\cite{Breuer_P}
\eq{BEq13}{
\partial_t\hat{\rho}=i[\hat{\rho},\tilde{H}_\Lambda]+\sum_{\alpha=1,2}\left(\mathcal{L}_{\alpha\rightarrow3}+\mathcal{L}_{3\rightarrow\alpha}\right)\hat{\rho}.
}
The second term describes the dissipative transitions via the superoperators $\mathcal{L}_{\alpha\rightarrow\beta}$, which act linearly on $\hat{\rho}$,
\eq{BEq14}{
\mathcal{L}_{\alpha\rightarrow\beta}\hat{\rho}=\gamma_{\alpha\rightarrow\beta}\left(|\beta\rangle\langle\alpha|\hat{\rho}|\alpha\rangle\langle\beta|-\frac{1}{2}\big\{|\alpha\rangle\langle\alpha|,\hat{\rho}\big\}\right).
}

\begin{figure}\center\includegraphics[width=1\linewidth]{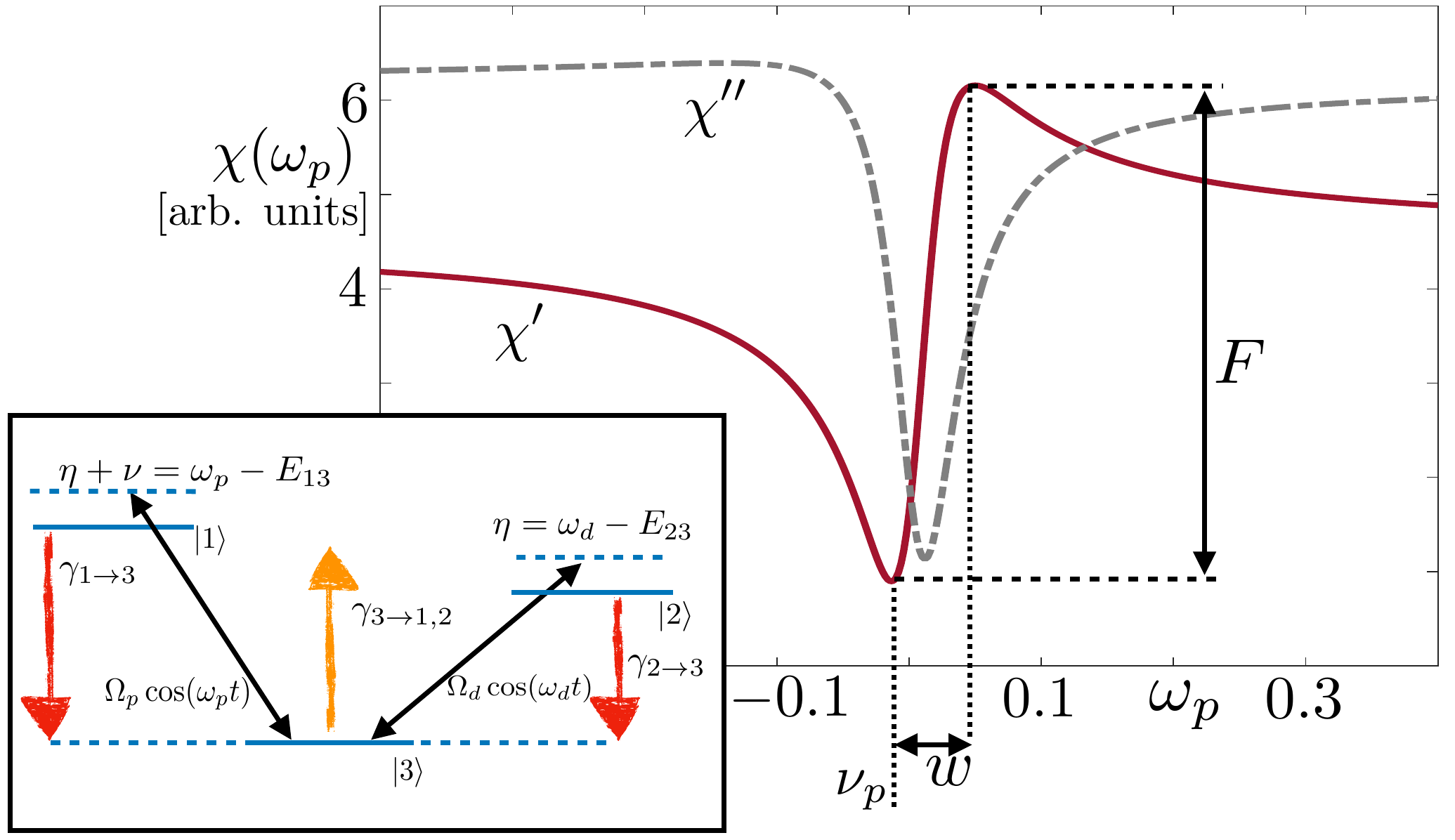}\caption{Spectral hole (Fano resonance) in the susceptibility $\chi(\omega_p)$ obtained in linear response in $\Omega_p$ from the $\Lambda$-scheme in Fig.~\ref{Fig3} (repeated in inset). Both the real ($\chi'$, red bold line) and the imaginary part ($\chi''$, grey dotted line) display an asymmetric line shape, indicating a Fano resonance close to the resonance condition $\omega_p-\omega_d=\approx E_{13}-E_{23}$. The dimensionless parameters for this figure are $\gamma=0.1, \Omega_d=4, \omega_d=18, E_{23}=20.4, E_{13}=19.4$. The dependence of the strength $F$, spectral width $w$ and position $\nu_p$ of the signal on the drive parameters can be found in Tab.~\ref{Tab1}. 
}
\label{Fig4}\end{figure}%%%

The linear response to the probe field $\sim\Omega_p$ is obtained from the stationary state ($\partial_t\hat{\rho}=0$) of Eq.~\eqref{BEq13}. To simplify notation, we assume one common rate $\gamma\equiv \gamma_{\alpha\rightarrow\beta}$ for all dissipative processes. This is justified for $k_BT >E_{13},E_{23}$. One finds
\eq{BEq11}{
\rho_{11}&\approx&\rho_{22}=\frac{1-\rho_{33}}{2},\\
\rho_{21}&=&\frac{\Omega_d\rho_{31}-\Omega_p\rho_{23}}{2(i\gamma-\nu)},\\
\rho_{23}&=&\frac{\Omega_d(3\rho_{33}-1)}{4\eta+6i\gamma},\\
\rho_{31}&=&\frac{i \gamma-\nu+\frac{\Omega_d^2}{4\eta+6 i \gamma}}{\Omega_d^2-(i\gamma-\nu) (6i \gamma -4 (\eta+\nu))}\Omega_p(3\rho_{22}-1).
}
The time-dependent expectation of an arbitrary, time-independent operator $\hat{O}$ in the rotating frame is  
\eq{BEq16}{
\langle\hat{O}\rangle(t)=\text{Tr}(U^\dagger(t)\hat{\rho} U(t)\hat{O})=\sum_{\alpha\beta}(U^\dagger(t)\hat{\rho} U(t))_{\alpha\beta}O_{\beta\alpha}.
}
If the response is evaluated at the probe frequency $\omega_p$, only terms proportional to $ \rho_{13}, \rho_{31}\sim e^{\pm i\omega_pt}$ contribute to Eq.~\eqref{BEq16}. This yields the linear response of the generic operator $\hat{O}$,
\eq{BEq17}{
\left.\frac{\partial\langle\hat{O}\rangle_{\omega_p}}{\partial \Omega_p}\right|_{\Omega_p=0}= \frac{O_{13}\left(i \gamma-\nu+\frac{\Omega_d^2}{4\eta+6 i \gamma}\right)}{\Omega_d^2-(i\gamma-\nu) (6i \gamma -4 (\eta+\nu))}(3\rho_{22}-1).\ \ \ \ \ \
}
For the specific choice of $\hat{O}=|1\rangle\langle3|+\text{H.c.}$, i.e., measuring the operator to which the probe field is coupled, $\frac{\partial\langle\hat{O}\rangle_{\omega_p}}{\partial \Omega_p}\equiv\chi_O(\omega_p)$ is the susceptibility.

The real and imaginary part $\chi'(\omega_d)=\text{Re}\chi(\omega_d)$ and $\chi''(\omega_d)\equiv \text{Im}\chi(\omega_d)$ of the susceptibility are shown in Fig.~\ref{Fig4} for a suitable set of parameters. They display a pronounced Fano resonance, i.e., a spectral hole, whose strength depends on the dissipation rate $\gamma$ and the Rabi frequency of the drive $\Omega_d$. For weak driving $\Omega_d\rightarrow 0$, the signal reduces to the expected Lorentzian $\chi''(\omega_p)= \frac{6\gamma}{36\gamma^2+16(\omega_p-E_{13})^2}$, with a peak $\sim\frac{1}{6\gamma}$ at $\omega_p=E_{13}$.

\subsection{Occurrence and strength of the Fano resonance }
The analytical form of the susceptibility $\chi(\omega_d)$ in Eq.~\eqref{BEq17} appears rather complicated. Especially when the contribution of several $\Lambda$-type schemes to the susceptibility is expected, the total signal is hard to estimate from the form of each individual $\chi(\omega_d)$ in Eq.~\eqref{BEq17}. In order to make the Fano signal more theoretically accessible, we perform a Taylor expansion of the imaginary part $\chi''(\omega_d)$ for strong and weak dissipation i.e., for $\gamma \gg \Omega_d, \eta$ and $\gamma\ll \Omega_d, \eta$. 

Experimentally and theoretically well accessible parameters, which characterize a Fano resonance are illustrated in Fig.~\ref{Fig4} and consist of the strength of the resonance $F$, its spectral width $w$ and its spectral peak position $\nu_p$. Their corresponding values obtained from Eq.~\eqref{BEq17} can be found in Table~\ref{Tab1}. In general, the Fano resonance is most significant for small dissipation rates $\gamma$ and small ratios $\frac{\eta}{\Omega_d}\ll 1$. The width and position of the signal can be adjusted by tuning $\eta$. 

Transferring the present analysis to the dimer and trimer level schemes shows that hole burning, i.e., Fano resonances in the magnetic susceptibility, can be observed in \Li already on the basis of magnetic dimers and trimers. One crucial requirement for its observation, however, are sufficiently small dissipation rates $\gamma$, which are of the order of the detunings $\nu, \eta$ and the Rabi frequency $\Omega_d$. According to Eq.~\eqref{BEq9} this can be achieved if the lattice degrees of freedom have a vanishing linewidth $\Gamma\rightarrow0$, i.e., in the limit of strongly isolated systems, which is in accordance with experimental findings\cite{Silevitch2017, Quilliam2007}.

\begin{table}[]
\begin{tabular}{|c|c|c|}\hline
parameter \ \ &\ \  weak $\gamma$ \ \   &\ \  strong $\gamma$  \\ \hline
$F$ & $\frac{1}{12\gamma}\frac{1}{1+\left(\frac{2\eta}{3\Omega_d}\right)^2}$ & $ \frac{1}{4\gamma}\left(\frac{\Omega_d}{3\gamma}\right)^2$  \\ \hline
 $w$& $\frac{2\eta}{3}\left(1+\frac{3\Omega_d^2}{8\eta^2}\right) $& $\frac{3\gamma^2}{2\eta}$   \\ \hline
 $\nu_p$& $4\eta \left(\frac{\Omega_d}{\eta}\right)^2$ &   $\frac{3\gamma^2}{2\eta}$ \\ \hline
\end{tabular}
\caption{Signal strength $F$, width $w$ and position $\nu_p$ of a Fano resonance in the $\Lambda$-scheme for weak and strong dissipation $\gamma$. Compare with Fig.~\ref{Fig4} for an illustration of the parameters.}\label{Tab1}
\end{table}

\section{Hole burning for many moments}\label{SecIII}
The mechanism which leads to the emergence of a Fano resonance in driven \Li can be understood in terms of a $\Lambda$-scheme in a Ho$^{3+}$ dimer configuration. A single dimer represents, however, an idealized setup, which completely neglects the many-body aspect of the magnetic moments in a \Li sample. In this section, we aim to generalize the previous findings to many interacting moments. Devising a phenomenological, effective spin-$\frac{1}{2}$ Hamiltonian, which models the low-energy Hilbert space of \Li {\it including} quantum corrections, we show that the driven system features an extensive number of many-body $\Lambda$-schemes, which can display Fano resonances under the above outlined conditions. Based on this finding, we argue that hole burning survives also in the realistic, many-body setting and that many-body $\Lambda$-schemes are in fact required in order observe Fano resonances at experimentally relevant conditions\cite{Silevitch2017}.

\subsection{Effective spin-$\frac{1}{2}$ quantum dipole Hamiltonian}
At low energies, the dynamics of the magnetic degrees of freedom in \Li is dominated by collective magnetic moments rather than by dimer or trimer configurations. In order to investigate the driving schemes in the many-body setup, we introduce an effective spin-$\frac{1}{2}$ quantum dipole Hamiltonian, which, on the one hand, is consistent with the results from the exact dimer and trimer analysis above and, on the other hand, recovers the common Ising approximation at large energies. 

In accordance with Eqs.~\eqref{BEq1} and \eqref{BEq2}, we propose a spin-$\frac{1}{2}$ Hamiltonian of the form
\eq{CEq1}{
H_{\frac{1}{2}}=A_{\text{dip}}C_{zz}^2 \sum_{l,m}\sum_{\alpha,\beta=x,y,z}\sigma^\alpha_l \sigma^\beta_m g^\alpha(\vec{R}_{lm})L^{\alpha\beta}(\vec{R}_{lm})g^\beta(\vec{R}_{lm}).\ \ 
}
Here $\sigma^\alpha_l$ is the Pauli matrix $\alpha=x,y,z$ describing the orientation of spin $l$ and $A_{\text{dip}}$, $C_{zz}$ and $L^{\alpha\beta}(\vec{R}_{lm})$ are the same as in Sec.~\ref{SecII}. The position dependent and dimensionless $g$-factors $g^{\alpha}(\vec{R}_{lm})$ are chosen such that $g^z(\vec{R}_{lm})=1$ and $g^{x,y}(\vec{R}_{lm})\ll1, \forall \vec{R}_{lm} $. In the limit $g^{x,y}(\vec{R}_{lm})\rightarrow0$, the Hamiltonian $H$ reduces to the Ising Hamiltonian in Eq.~\eqref{BEq2}. 

We introduce non-zero $g^{x,y}(\vec{R}_{lm})$ to describe deviations of the true \Li system from the Ising approximation. For the dimer configuration, this deviation vanishes at large distances with $\sim |\vec{R}_{lm}|^{-3}$, as discussed in Eq.~\eqref{BEq4}. This suggests $g^{x,y}(\vec{R}_{lm})\sim |\vec{R}_{lm}|^{-3}$ in a similar fashion. For more than two magnetic moments (e.g. in trimer configurations) we find, however, that non-Ising corrections decay much slower in the distance $|\vec{R}_{lm}|$. We attribute this behavior to strong contributions to the dipole-dipole interactions from highly excited crystal field eigenstates, which were observed in perturbation theory for the dimer setup. We expect the non-Ising corrections therefore to become more pronounced for larger spin clusters and thus chose $g^{x,y}(\vec{R}_{lm})\equiv g^{x,y}$ independent of the distance. A similar effective model has also been proposed to explain temperature dependence of the specific heat of \Li~\cite{Ghosh2003}.

For the choice $g^x=0.1, g^y=0.07$, the eigenvalues of $H_{\frac{1}{2}}$ match well with the behavior of the lowest order eigenvalues of the full Hamiltonian $H$ for $n=2,3$ spins. Compared to previous approaches\cite{Ghosh2003}, the $g$-factors here are not isotropic. This is a necessary requirement in order to obtain the observed level splittings and a $\Lambda$-type driving scheme. This is consistent with the anisotropy of the crystal field Hamiltonian\cite{Ronnow2007}.

In order to simulate a realistic subsystem of \Li with $x\le0.05$, we consider a three-dimensional volume of $N=5\times 5\times 3$ unit cells and we randomly distribute $n=12$ spins over the $4\times N=300$ potential Ho$^{3+}$ positions. This corresponds to a dilution of $x=0.04$. The spins experience dipole-dipole interactions, which are described by the Hamiltonian $H_{\frac{1}{2}}$ in Eq.~\eqref{CEq1}. The eigenvalues and eigenstates of the many-body spin Hamiltonian are obtained via exact diagonalization.  

The sequence of eigenvalues can be understood in a similar way as for the trimer scheme. Consider the eigenvalues $\lambda_l$ of $H_{\frac{1}{2}}$ with $l=1, ..., 2^{12}$ and sorted in ascending order, i.e. 
\begin{figure}[H]
\center
\includegraphics[width=1\linewidth]{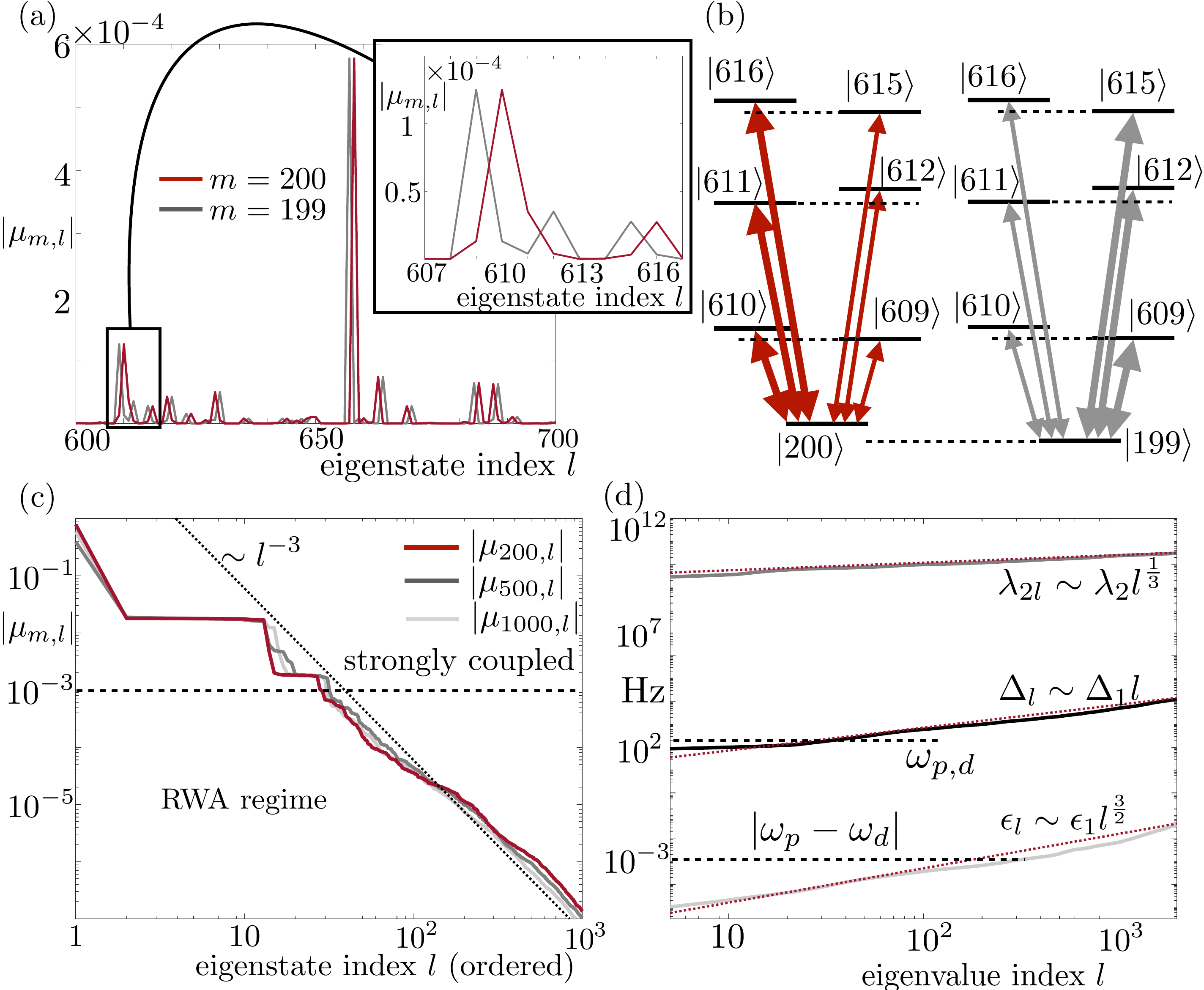}
\caption{The energy spectrum and the eigenstates of the effective Hamiltonian $H_{\frac{1}{2}}$ in Eq.~\eqref{CEq1} confirm the observation from the dimer and trimer configurations: Due to the weakly broken Ising symmetry in $H_{\frac{1}{2}}$ its eigenstates come in pairs, each consisting of two quasi-degenerate states with relative level splitting $\epsilon$, and separated from other pairs by an "Ising" energy $\Delta$.  For eigenvalues $\lambda_l$ of $H_{\frac{1}{2}}$ in ascending order, we define the quasi-degeneracies $\epsilon_{l}\equiv\lambda{2l}-\lambda_{2l-1}$ and the "Ising" energies $\Delta_{l}=\lambda_{2l+1}-\lambda_{2l}$. These are shown in panel (d). For a cluster of $n=12$ random magnetic moments, corresponding to a small \Li crystal with $x=0.04$, one finds two well separated energy bands of quasi-degeneracies and Ising energies. The energies of the Ising band and the quasi-degenerate band of the cluster correspond well with the drive, probe frequency $\omega_{p,d}\approx 2\pi\times200$Hz and their relative detuning $\delta\omega=\omega_p-\omega_d\sim 2$mHz used in previous hole burning experiments\cite{Silevitch2017}. (a+b) The transition matrix elements $\mu_{m,l}=\langle m|J^z_{\text{tot}}|l\rangle$ between different eigenstates $|l\rangle, |m\rangle$ establish a set of $\Lambda$-schemes, similar to the dimer and trimer configurations. In (a) this is shown for the absolute values of $\mu_{m,l}$ for the quasi-degenerate pair $m=199, 200$ and for $500<l<700$. The pairwise $\Lambda$-scheme structure is illustrated in (b) for the states in the inset.  Thick arrows correspond to large matrix elements and thin arrows to small matrix elements.  The whole set of coupling matrix elements for a given eigenstate $m=200,500,1000$ from the $n=12$ spin cluster is shown in (c), where the $|\mu_{m,l}|$ are sorted in descending order. Matrix elements $|\mu_{l,m}|<10^{-3}$ correspond to Rabi frequencies $\Omega_d=O($kHz$)$ in the presence of a $h_d=0.5$Oe drive field\cite{Silevitch}. For drive and probe frequencies $\omega_{d,p}\approx 2\pi\times 200$Hz, this corresponds to weak driving $\Omega_{d,p}\le\omega_{d,p}$ and allows us to treat the response of the system in the rotating wave approximation (RWA). The minority of strongly coupled transitions has no observable influence on the dynamics.  In this framework, hole burning, i.e., a Fano resonance in the linear magnetic susceptibility, is observable when the detuning $\delta\omega$ equals the energy $\epsilon_l$ of one (or several) quasi-degeneracies. This leads to destructive interference between two different pathways in the $\Lambda$-scheme and to a Fano resonance as in Fig.~\ref{Fig4}.
}\label{Fig6}\end{figure}%

$\lambda_{l+1}>\lambda_l$ for all $l$. We define the 'Ising' level spacings $\Delta_j$ and the 'quantum' level spacings $\epsilon_j$ according to
\eq{CEq2}{
\epsilon_j&=& \lambda_{2j}-\lambda_{2j-1}, \text{ for } j=1, ... 2^{12},\\
\Delta_j&=&\lambda_{2j+1}-\lambda_{2j}, \text { for } j=1,... 2^{12}-1.
} 
The level spacings $\Delta_j, \epsilon_j$ are both positive for all $j$ and the values of the $\epsilon_j$'s are a measure for the deviation of $H_{\frac{1}{2}}$ from $H_{\text{Ising}}$. For $g^{x,y}\rightarrow0$, $\epsilon_j\rightarrow0$ continuously as $H_{\frac{1}{2}}$ approaches $H_{\text{Ising}}$. A characteristic distribution of $\{\lambda_{2j}, \Delta_j, \epsilon_j\}$ for a system of $n=12$ spins is shown in Fig.~\ref{Fig6} (d).

The importance of the level spacings $\epsilon_j$ becomes apparent when the many-spin system is driven with a time dependent magnetic field in the $z$-direction. As in the previous section, this is formally described by adding a Hamiltonian $H\rightarrow H(t)=H_{\frac{1}{2}}+\delta H(t)$ with $\delta H(t)=h_d\cos(\omega_dt)C_{zz}g_L\mu_B\mu_0 S^z_{\text{tot}}$ and $S^z_{\text{tot}}=\sum_{l=1}^{12}\sigma^z_l$. The transition matrix elements between two eigenstates $|l\rangle, |m\rangle$ of the Hamiltonian $H_{\frac{1}{2}}$ are
\eq{CEq3}{
\mu_{l,m}=\langle l| S^z_{\text{tot}}|m\rangle.
}
As in the previous dimer and trimer configurations, $[S^z_{\text{tot}},H_{\text{Ising}}]=0$ but $[S^z_{\text{tot}},H_{\frac{1}{2}}]\neq0$, and thus the transition matrix can be chosen diagonal in the Ising basis but will be non-diagonal in the basis of $H_{\frac{1}{2}}$. 

Figure~\ref{Fig6} (a) shows the matrix elements $\mu_{m,l}$ for fixed $m=199$ (red line) and $m=200$ (grey line). One observes a generalization of the dimer and trimer drive schemes to the many-spin system. Both states $m=199, 200$ act as the base state of a whole set of inverse $\Lambda$-schemes, which couple to pairs of states $l=2j-1, 2j$. The levels of each pair are separated by a 'quantum' level spacing $\epsilon_j$ (see Fig.~\ref{Fig6} (b) for an illustration). Each $\Lambda$-scheme consists of one strong and one weak transition matrix element, i.e. one generally finds either $|\mu_{199,l}| \ll |\mu_{199,l+1}|$ or $|\mu_{199,l}| \gg |\mu_{199,l+1}|$. This order is exchanged when going from $m$ to $m+1$, as can be seen from the inset of Fig.~\ref{Fig6}~(a) and the arrows in Fig.~\ref{Fig6}~(b). The pairs, in turn, are separated from the base state $m=199,200$ by one or several Ising level spacings $\Delta_m$. 

The complete set of transition matrix elements $\mu_{m,l}$ for a selection of fixed $m$ is plotted in Fig.~\ref{Fig6}~(c) in descending order. One observes a small number of about $10$ matrix elements for each $m$, which are $\sim O(1)$. For $l\ge 10$ one observes a significant drop in the magnitude of $|\mu_{m,l}|$, which is followed by a decay  $\sim l^{-3}$. While for each $m$ the states corresponding to a given $l$ are different (due to individual ordering) the magnitude and decay of the matrix elements is very similar. 

Large matrix elements $|\mu_{m,l}|=O(1)$ result from overlaps of nearly Ising or $\mathds{Z}_2$-reversed partners. Consider therefore a state $\alpha$ of the particular form $|\Uparrow_\alpha\rangle\equiv |\uparrow\uparrow\downarrow\uparrow...\rangle$ and its $\mathds{Z}_2$-reversed partner $|\Downarrow_\alpha\rangle\equiv|\downarrow\downarrow\uparrow\downarrow...\rangle$, where $\alpha$ is the label that indicates which spins are pointing up and which ones are pointing down. Due to the smallness of $g^{x,y}$, many eigenstates will be of the form $|\psi_\alpha\pm\rangle=\frac{|\Uparrow_\alpha\rangle\pm|\Downarrow_\alpha\rangle}{\sqrt{2}}+...$, where $...$ indicates perturbative corrections due to non-zero $g^{x,y}$. The largest transition matrix elements result from overlaps $\langle \psi_\alpha +|S^z_{\text{tot}}|\psi_\alpha -\rangle=n_{\uparrow}-n_{\downarrow}+...$, which is the difference in the number of up-spins and down-spins and is $O(1)$. All the remaining matrix elements with $|\mu_{m,l}|\ll1$ are due to the corrections $...$ and, as we will see, dominate the dynamics under driving. 

\subsection{Driving the many-spin system}
In this section, we discuss the response of the many-spin system with its multiple $\Lambda$-schemes to external driving. 

\begin{figure}
\center\includegraphics[width=\linewidth]{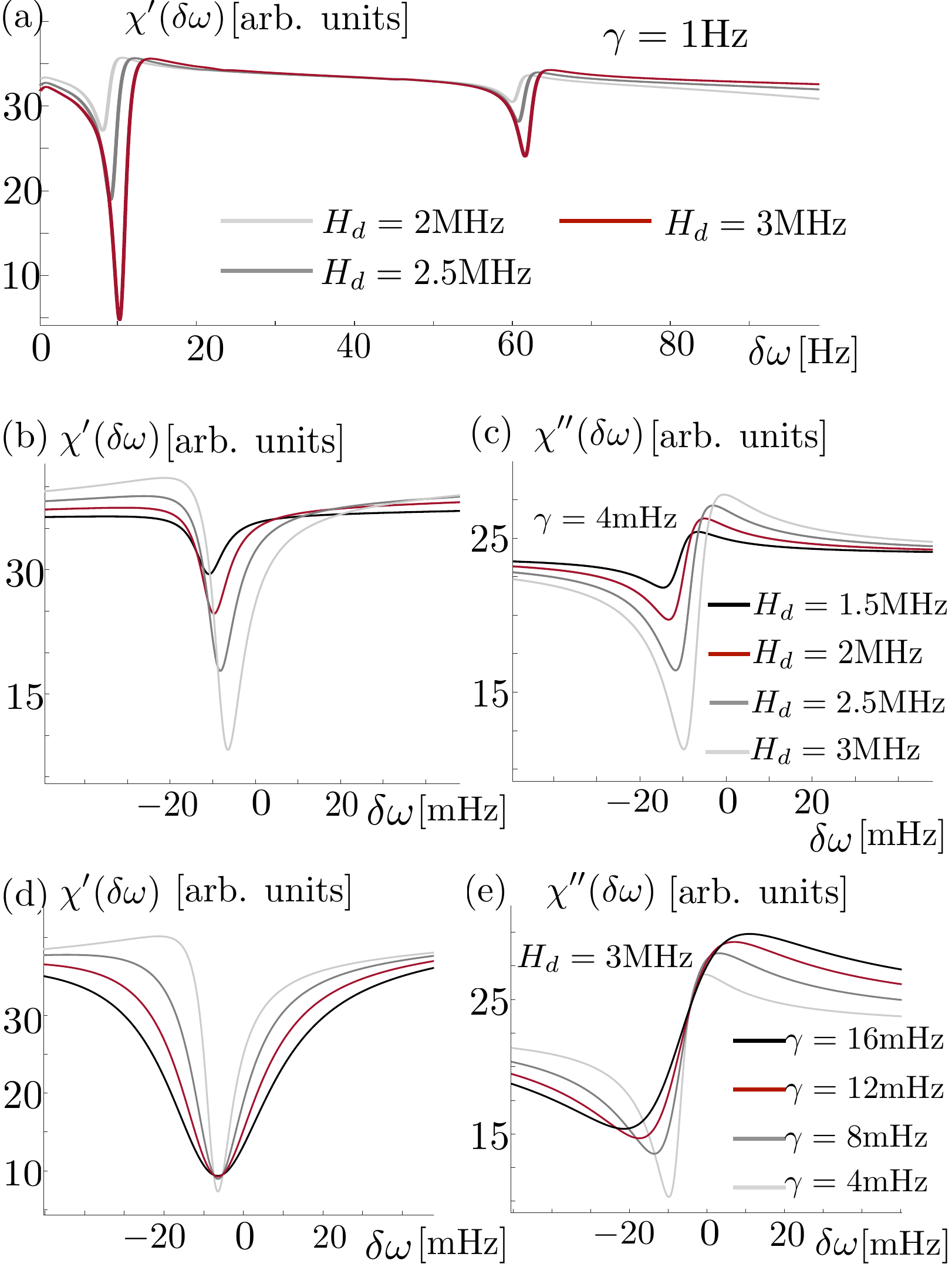}\caption{For suitable driving conditions, the combination of quasi-degenerate and Ising energy levels with the $\Lambda$-schemes in $\mu_{l,m}$ cause observable Fano resonances in the linear magnetic susceptibility $\chi(\omega)$. The plots (a-e) show Fano resonances in the real part $\chi'$ and the imaginary part $\chi''$ of the susceptibility for different dissipation rates $\gamma$. The signal is obtained from $n=12$ magnetic moments, which are described by $H_{\frac{1}{2}}$ and driven by a magnetic field with frequency $\omega_d=2\pi\times202$Hz and variable strength $H_d$. $\chi$ is probed at frequency $\omega_p=\delta\omega+\omega_d$. The drive strength $H_d=3$MHz corresponds to a magnetic field of $h_d\approx0.2$Oe. In order to obtain an observable signal, the width of the resonance, i.e., the quasi-degeneracy, has to match approximately with both the phonon induced dissipation rate $\gamma$ and the detuning $\delta\omega$ and, in addition, the Rabi frequency $\Omega_d$ for the transition needs to be sufficiently large, $|\Omega_d|\gg\delta\omega$, to cause interference. If $\gamma$ or $H_d$ are changed considerably a given resonance vanishes and the signal becomes flat until another resonance becomes accessible. 
}\label{Figfin}\end{figure}
The driving regime of interest is the one discussed in Refs.~\cite{Ghosh2002a,Silevitch2007a,Silevitch2017}, where a clear Fano resonance has been observed. The setup consists of a \Li sample, which is driven by two different, time-dependent magnetic fields, a driving field $\sim h_d\cos(\omega_d t)$ and a probe field $\sim h_p\cos(\omega_p t)$ with small amplitude $h_p\ll h_d$. Typical experimental values for the drive and the probe frequency are $\omega_{d,p}\approx 2\pi\times 200$Hz$\approx 1.2$kHz and for their difference $\delta\omega=|\omega_{p}-\omega_{d}|\le 2\pi\times 10$mHz. The strength of both the pump and the probe field is $h_{d}\approx 0.5$Oe and $h_{p}=0.02$Oe. For this choice, the corresponding Rabi frequencies $\Omega_{l,m}$ for transitions between spin eigenstates $|l\rangle\leftrightarrow|m\rangle$ are 
\eq{CEq4}{
\Omega_{l,m}^{d,p}=\underbrace{h_{d,p}g_L C_{zz} \mu_{B}\mu_0}_{\equiv H_{d,p}}\mu_{l,m},
}
which amounts to $\Omega_{l,m}^{d}\approx \mu_{l,m}\times9.5\text{MHz}$ and $\Omega^p_{l,m}=\frac{\Omega_{l,m}^{d}}{20}$. We defined the effective driving, probing field $H_{d,p}$ for brevity. 

We distinguish two different regimes for the Rabi frequencies $\Omega_{l,m}^{d,p}$ and the matrix elements $\mu_{l,m}$: (i) a regime of strong driving with $\Omega_{l,m}^{d,p}>\omega_{d,p}$ and (ii) a rotating wave regime (RWA) for $\Omega_{l,m}^{d,p}<\omega_{d,p}$. By definition, the conditions for the strong coupling regime deny the application of the rotating wave approximation and the corresponding transitions have to be treated in the Floquet formalism\cite{Nori2007,Grifoni2010}. In the RWA regime, however, the rotating wave approximation is applicable and the discussion of Sec.~\ref{SecII} can be generalized to the multi-spin setup. 

For the above mentioned parameters\cite{Ghosh2002a,Silevitch2007a,Silevitch2017}, the two regimes are illustrated in Fig.~\ref{Fig6} (c). It shows that this particular choice of driving parameters leads to a clear separation between the strong driving regime and the RWA regime, which is indicated by a jump of $\mu_{l,m}$ over at least one order of magnitude after escaping the RWA regime and only a few matrix elements that exceed slightly the RWA condition. This is further justification why we can treat the strong coupling and RWA regime separately. In Appendix~\ref{StrongDr} we show that very strongly driven transitions $\Omega^d\gg\omega^d$ will effectively freeze out and need not be considered. We will thus focus on the RWA regime. 

In the RWA regime, the analysis of Sec.~\ref{SecII} can be generalized almost straightforwardly to the case of many $\Lambda$-schemes. One difference between the idealized scheme and the real driving scheme is, however, that both the drive and the probe field couple to the same transition matrix elements. This yields the time-dependent Hamiltonian
\eq{Extra3}{
H(t)=H_{\frac{1}{2}}+\Big(H_d\cos(\omega_dt)+H_p\cos(\omega_pt)\Big)\sum_l \sigma^z_l.
}
Considering a single $\Lambda$-scheme $|l\rangle\leftrightarrow|m\rangle\leftrightarrow|l+1\rangle$, both transitions couple to the combined magnetic field, which gives rise to two meaningful ways of going to a rotating frame. One is obtained by performing the rotating transformation as in Eq.~\eqref{BEq11} and yields 
\eq{Extra4}{
H_{\text{RWA}}&=&\left(\frac{H_d+H_p e^{i\delta\omega t}}{2}\mu_{l,m}|l\rangle\langle m|+\text{H.c.}\right)\\
&&+\left(\frac{H_p+H_d e^{-i\delta\omega t}}{2}\mu_{l+1,m}|l+1\rangle\langle m|+\text{H.c.}\right)\nonumber\\
&&+\nu | l \rangle\langle l|+(\nu+\eta)|m\rangle\langle m|\nonumber.
}
It still contains slowly varying terms with frequency $\delta\omega=\omega_p-\omega_d$. A similar transformation is obtained by exchanging $l\leftrightarrow l+1$ in the transformation matrix which yields $H_{\text{RWA}}$ but with $H_d\leftrightarrow H_p$ and $\nu\rightarrow2\delta\omega-\nu, \eta\rightarrow \eta+\nu-\delta\omega$. Both Hamiltonians yield the equivalent time evolution since in both transformations only the fast contributions $\sim 2\omega_p, 2\omega_d$ and $\omega_p+\omega_d$ have been neglected. The ambiguity in choosing the transformation reflects the fact that, when measuring at the frequency $\omega_p$, one can  either probe the $l\leftrightarrow m$ transition (corresponding to Eq.~\eqref{Extra4}) or the $l+1\leftrightarrow m$ transition corresponding to the second transformation. Per the $\Lambda$-scheme, one can thus probe two different transitions, which we take into account individually.

For the many-body scheme in Fig.~\ref{Fig6} (b), the magnetic susceptibility at the probe frequency $\chi(\omega_p)$ is given by the sum of all possible transitions, i.e., by the sum over all $\Lambda$-schemes with two different contributions per scheme. At extremely long measurement times $\sim \frac{2}{\delta\omega}$ the experiment\cite{Silevitch2017, Schmidt2014} singles out contributions at $\omega_p$ and discards all other parts. In linear response, the dimensionless susceptibility is $\chi(\omega_p)=\left. \frac{\partial\langle S^z_{\text{tot}}\rangle_{\omega_p}}{\partial H_p}\right|_{H_p=0}$. For a single $\Lambda$-scheme from Fig.~\ref{Fig6}~(b) of the form  $|l\rangle\leftrightarrow |m\rangle\leftrightarrow |l+1\rangle$ it acquires two contributions, one probing the $|l\rangle \leftrightarrow |m\rangle$-transition and one probing the $|l+1\rangle\leftrightarrow|m\rangle$-transition, which yields
\eq{Eqss}{
\tilde\chi_{l,m}(\omega_p)&=&\frac{|\mu_{l,m}|^2\left(i\gamma-\nu+\frac{H_d^2|\mu_{l+1,m}|^2}{4\eta+6i\gamma}\right)}{H_d^2|\mu_{l+1,m}|^2-(i\gamma-\nu)(6i\gamma-4(\eta+\nu))}\\
&&+\frac{|\mu_{l+1,m}|^2\left(i\gamma+\nu-2\delta\omega+\frac{H_d^2|\mu_{l,m}|^2}{4(\eta+\nu+\delta\omega)+6i\gamma}\right)}{H_d^2|\mu_{l,m}|^2-(i\gamma+\nu-2\delta\omega)(6i\gamma-4(\eta+\delta\omega))}.\nonumber
}
Here, according to the definitions in Sec.~\ref{SecIIB},  $\nu=\lambda_{l+1}-\lambda_l+\omega_d-\omega_p$ and $\eta=\lambda_l-\lambda_m-\omega_d$. The signal corresponding to the smaller transition matrix element is strongly suppressed compared to the one corresponding to the larger matrix element due to the $|\mu|^2$ prefactor and rarely contributes to the susceptibility. As a consequence, the total susceptibility $\chi(\omega_p)=\sum_{l,m}\tilde\chi_{2l-1,m}(\omega_p)$ is very well approximated as the sum of independent $\Lambda$-schemes.

In order to contribute a Fano resonance to the magnetic response, a given $\Lambda$-scheme has to produce a significant signal strength $F$ at a small signal width $w$, as shown in Tab.~\ref{Tab1}. For a transition $|l\rangle\leftrightarrow|m\rangle\leftrightarrow |l+1\rangle$ to contribute this requires a near resonant detuning from the drive frequency $\eta=\omega_d-|\lambda_l-\lambda_m|=O(\text{mHz})$ and in addition a transition matrix element $\mu_{l,m}$ in the RWA regime and a detuning of the probe frequency $\delta=\epsilon_{l/2}- |\omega_d-\omega_p|=O(\text{mHz})$.

 \begin{figure}[h]\center\includegraphics[width=\linewidth]{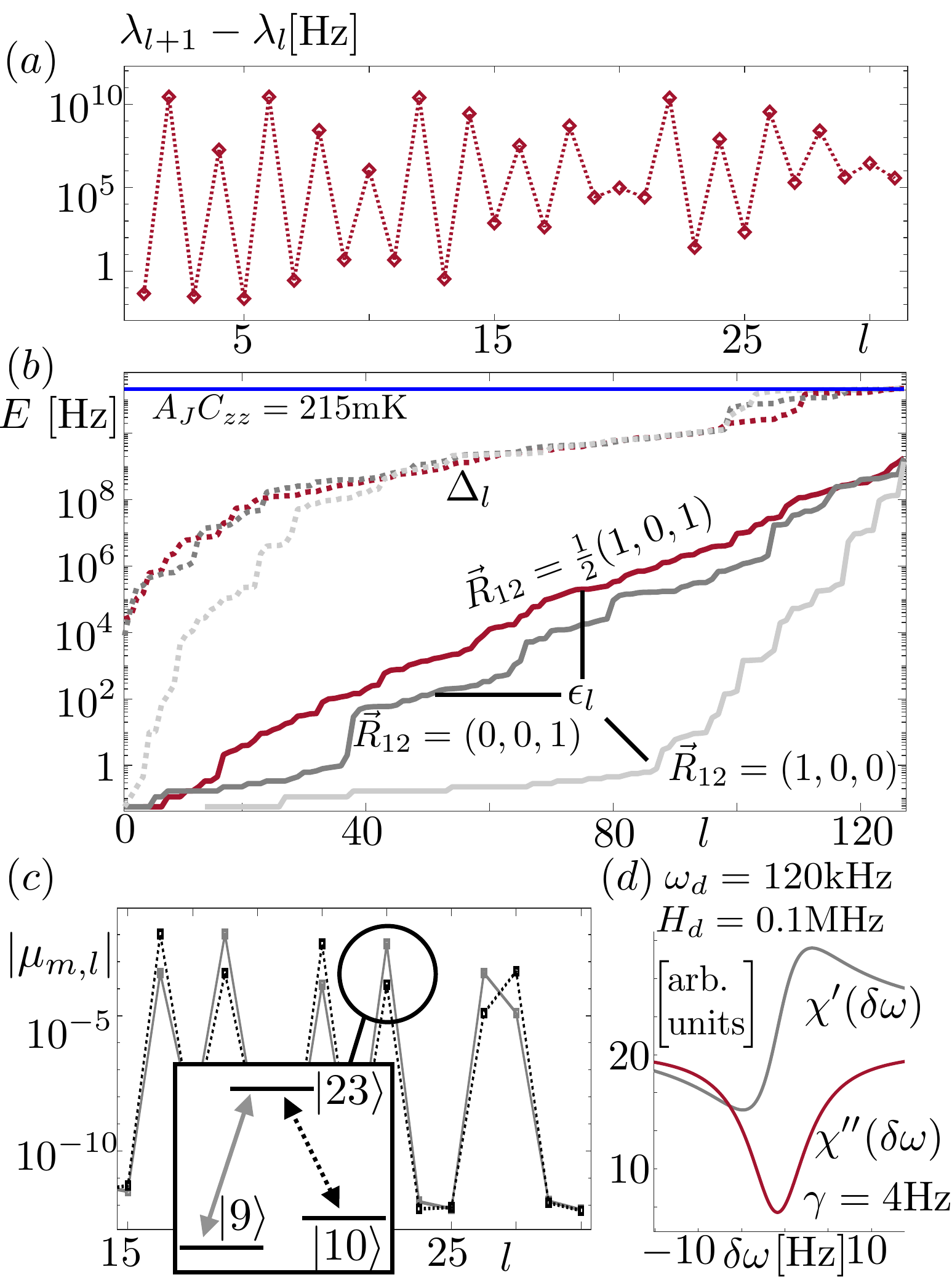}\caption{The hole burning phenomenology persists after including hyperfine interactions between the magnetic moments and the Ho nuclear spins. Including the nuclear spin degrees of freedom in a magnetic \Li dimer configuration with relative orientation $\vec{R}_{12}$ confirms the established picture of quasi-degenerate pairs of eigenstates of the Hamiltonian $H^{(2)}_{\text{full}}$ in Eq.~\eqref{Hf3} and $\Lambda$-schemes in their transition matrix elements. The alternating level structure of Ising-type level differences $\Delta$ (even $l$) and quasi-degeneracies $\epsilon$ (odd $l$) can be found throughout the entire spectrum of $H^{(2)}_{\text{full}}$ and is illustrated in (a) for the $30$ lowest energy eigenstates. This leads to energetically well separated bands of level differences $\Delta_l=\lambda_{2l+1}-\lambda_{2l}$ and quasi-degeneracies $\epsilon_l=\lambda_{2l}-\lambda_{2l-1}$, shown in (b) for different dimer orientations $\vec{R}_{12}$, as it was observed for magnetic clusters without hyperfine interactions. The transition matrix elements $\mu_{m,l}=\langle m| J^z_1+J^z_2|l\rangle$ between different dimer eigenstates $|m, l\rangle$ reveal multiple $\Lambda$-schemes. For the quasi-degenerate pair $m=9,10$ and $15<l<30$, this is demonstrated in (c), where the $|9\rangle\leftrightarrow|24\rangle\leftrightarrow|10\rangle$ transition is highlighted in the inset. This combination of level spacings and transition matrix elements again enables Fano resonances in the linear susceptibility, which is demonstrated in (d) for a driven \Li dimer for a specific set of drive parameters. %Quantitatively, the hyperfine interaction increases the Ising level splitting and suppresses quantum corrections from the dipole-dipole interaction. This shifts both $\mu_{m,l}$ and $\epsilon_l$ to smaller absolute values.
}\label{HFPlot}\end{figure}%%%

In Fig.~\ref{Figfin}, we show the magnetic susceptibility $\chi$ of a system of $n=12$ magnetic moments, which is described by $H_{\frac{1}{2}}$ with realistic parameters for \Li. It is strongly driven by an external drive field $H_d=1.5-3$MHz, which corresponds to $h_d\approx 0.1-0.2$Oe. We clearly observe pronounced, individual Fano resonances as a function of the probe field detuning $\delta\omega=\omega_p-\omega_d$ whose emergence and visibility depend on the dissipation rate $\gamma$, the drive field strength $H_d$ and the drive frequency $\omega_d$, as predicted by the $\Lambda$-scheme analysis in Sec.~\ref{SecII}. The observed resonances correspond to $\Lambda$-schemes for which the detuning $\eta$, the Rabi frequency $H_d\mu_{l,m}$ and the dissipation are roughly of the same order of magnitude (mHz for Fig.~\ref{Figfin}(b-e) and Hz for Fig~\ref{Figfin}(a)). 

\subsection{The effect of hyperfine interactions\label{nuke}}
The spin$-\frac{1}{2}$ toy model in Eq.~\eqref{CEq1} predicts the observation of hole burning at quantitatively correct energy scales in \Li under experimentally realistic conditions. The shape of $H_{\frac{1}{2}}$ in Eq.~\eqref{CEq1} is motivated by the microscopic Hamiltonian in Eq.~\eqref{Eq1}, which predicted hole burning by the same mechanism as for $H_{\frac{1}{2}}$ but only for very different energy scales, which correspond to the flipping of a single magnetic moment. Throughout this discussion, we have completely neglected the hyperfine interaction of the electron magnetic moment $\vec{J}$ with the nuclear moments $\vec{I}$ of the Ho atoms. The effect of hyperfine interactions in \Li has been addressed by several papers\cite{Schechter2005,Schechter2008,Bitko1996,Chin2006} and, in accordance with their findings, we argue that the hyperfine interactions do not modify our hole burning phenomenology for sufficiently small transverse magnetic fields.  

The microscopic hyperfine interaction is described by the Hamiltonian 
\eq{Hf}{
H_{\text{hf}}=A_J \sum_l\left[I^z_lJ^z_l+\frac{1}{2}\left(I^+_lJ^-_l+I^-_lJ^+_l\right)\right]}
with $A_J=39$ mK and a nuclear spin $I=\frac{7}{2}$. The longitudinal part $\sim A_JJ^z_lI^z_l$ splits each electronic angular momentum state into a multiplet with eight nuclear spin states $m_J=-\frac{7}{2},...\frac{7}{2}$. In the Ising approximation, the hyperfine interaction reduces to
\eq{Hf2}{
H_{\text{hf-Ising}}=A_JC_{zz} \sum_l \sigma^z_lJ^z_l
}
and each state is separated from its adjacent states $m_J\pm1$ by the energy $\sim A_JC_{zz}=215$ mK.

Both $H_{\text{hf}}$ and $H_{\text{hf-Ising}}$ are invariant under $(J^z_l, I^z_l) \rightarrow (-J^z_l, -I^z_l)$ or $(\sigma^z_l, I^z_l) \rightarrow (-\sigma^z_l, -I^z_l)$, respectively, and thus respect the Ising symmetry of the ground state manifold of the crystal field Hamiltonian. The leading order corrections to the Ising approximation are thus again arising from the dipole-dipole interactions between electronic magnetic moments of the form $\sim J_l^xJ_m^z$ (or $\sim \sigma^x_l\sigma^z_m$ in $H_{\frac{1}{2}}$). Compared to the situation without nuclear moments one, however, expects the hyperfine interactions to further suppress the corresponding quantum corrections. Qualitatively, this is due to the energy cost associated with changing the orientation of the electronic spin by applying $J^x_l$ to the electronic magnetic moment while at the same time leaving the nuclear spin orientation unchanged. 

At low temperatures $T=O(0.1\text{K})$, the excited states of the crystal field Hamiltonian remain inaccessible (apart from virtual excitations) due to their large energy separation $\sim 10.5$K. This remains true in the presence of hyperfine interactions. The symmetry breaking terms $\sim J_l^xJ_m^z$ again induce transitions only inside the ground state manifold of the crystal field Hamiltonian $|\uparrow\rangle_l\leftrightarrow|\downarrow\rangle$, which now experiences an additional energetic suppression given by the difference in the hyperfine interaction energy $\Delta E_{\text{hf}}$. According to Eq.~\eqref{Hf2} it is approximately $\Delta E_{\text{hf}}\approx 2A_JC_{zz} |m_J|$ and for a given electronic magnetic moment $l$ the additional suppression of quantum corrections may be expected to be proportional to its nuclear spin orientation. 

As a consequence, the phenomenology of hole burning arising from Ising symmetry breaking dipole-dipole interactions, which lift the degeneracy between Ising-reversed partners and introduce small but finite transition matrix elements $\mu_{\alpha\beta}=\langle \alpha|\sum_l J^z_l|\beta\rangle$ would survive, with quantitative corrections, also in the presence of hyperfine interactions. In order to test this assumption, we determine the eigenenergies and eigenstates of a microscopic dimer configuration in \Li including hyperfine interactions. Each electronic magnetic moment $\vec{J}_{1,2}$ then experiences the crystal field, the nuclear spin of the Ho$^{3+}$ ion and the mutual magnetic dipole-dipole interaction. This extends the dimer Hamiltonian $H^{(2)}$ in Eq.~\eqref{BEq1} to
\eq{Hf3}{
H^{(2)}_{\text{full}}=H^{(2)}+A_J\sum_{l=1}^2\vec{J}_l\cdot\vec{I}_l.
}
We diagonalize this Hamiltonian numerically in the  $(17\times8)^2$-dimensional Hilbert space and then inspect the $(2\times8)^2$-dimensional subspace of low energy eigenstates. 

The results obtained from the diagonalization confirm the above picture and support our phenomenology of hole burning. As for the dimer and trimer schemes without hyperfine interactions, each eigenstate of $H^{(2)}_{\text{full}}$ comes with a quasi-degenerate partner. In Fig.~\ref{HFPlot} this is demonstrated for a dimer with relative orientation $\vec{R}_{12}=(a/2,0,c/4)$ where $a,c$ are the \Li lattice constants. Figure \ref{HFPlot}(a) shows the level differences $\lambda_{l+1}-\lambda_l$ for the $m=32$ lowest energy states in the dimer, whose alternating pattern reveals the quasi-degeneracies. This represents an extension of the dimer energy levels without nuclear spins shown in Fig.~\ref{Fig1} (b,d). 

Grouping the differences of adjacent energies in the dimer spectrum into quasi-degenerate level splittings $\epsilon_l\equiv \lambda_{2l}-\lambda_{2l-1}$ and "Ising"-splittings $\Delta_l=\lambda_{2l+1}-\lambda_{2l}$, each dimer configuration now gives rise to a whole band of splittings, shown in Fig.~\ref{HFPlot} (b) for different configurations $\vec{R}_{12}$. Compared withFig.~\ref{Fig1}, the hyperfine interactions generally suppress both $\epsilon_l$ and $\Delta_l$. A trend towards stronger suppression for an increasing polarization of the nuclear moments, $P^{(2)}_z=|\langle I^z_1\rangle|+|\langle I^z_2\rangle|$, in a given set of quasi-degenerate states is observed.

As for the dimer setup without nuclear spins, the degeneracy breaking dipole-dipole interactions induce non-zero transition matrix elements $\mu_{\alpha,\beta}\equiv \langle \alpha|\sum_{l}J^z_l |\beta\rangle$ between different eigenstates $|\alpha,\beta\rangle$ of $H^{(2)}_{\text{full}}$. In Fig.~\ref{HFPlot} (c), these are shown for fixed $\alpha=9,10$ (two quasi-degenerate partners from the spectrum) and $\beta=15,...,30$ for the same configuration as in (a). It implies that in the presence of a time dependent external magnetic field $\sim \sum_{l}J^z_l$ two quasi-degenerate partners build out several $\Lambda$-schemes with alternating strong and weak transitions, very similar to the scheme observed in Fig.~\ref{Fig6} (b) for clusters of magnetic moments without hyperfine interactions. 

In the presence of hyperfine interactions, a single dimer scheme thus already contributes a whole set of many-body $\Lambda$-schemes, which can support Fano resonances over a much larger frequency range than a dimer scheme without hyperfine interactions. For example, the magnetic susceptibility $\chi(\omega)$ for the dimer configuration with $\vec{R}_{12}=(a/2,0,c/4)$ is shown in Fig.~\ref{HFPlot} (d) for a dissipation rate $\gamma=4$Hz, a drive field amplitude $h_d=10$mOe and frequencies $\omega_d=120$kHz, $\omega_p=\omega_d+\delta\omega$, with the use of Eq.~\eqref{Eqss} and by summing over the $m=256$ low energy eigenstates.

In conclusion, the consideration of hyperfine interactions extends the $\Lambda$-scheme of a dimer configuration of Ho$^{3+}$ magnetic moments to several, many-body $\Lambda$-schemes, each of which has the potential to establish a Fano resonance in the magnetic susceptibility when driving the system with a strong external magnetic field. Although the hyperfine interactions suppress the quasi-degenerate splittings $\epsilon_\alpha$ and the corresponding transition matrix elements $\mu_{\alpha,\beta}$ in the dimer, their numerical values are still larger than what is observed in experimental measurements \cite{Silevitch2017,Schmidt2014}. This indicates that the true, experimentally observed hole burning actually results from an interplay of dipole-dipole interactions between many electronic magnetic moments on the one hand and hyperfine interactions on the other hand. The basic phenomenology remains the same in the presence of hyperfine interactions, but they suppress quantum effects, which effectively decreases the necessary size of magnetic clusters in order to observe Fano resonances at the millihertz scale.

\subsection{Inferring dissipation scales from experimental data}
The analysis of the effective Hamiltonian $H_{\frac{1}{2}}$ motivated the assumption of isolated, many-body $\Lambda$-schemes, for which Eq.~\eqref{Eqss} is applicable and shows low-energy Fano resonances in the magnetic susceptibility comparable with the experimentally observed amplitude and frequency scales. In order to observe resonances, we had, however, to {\it guess} a suitable value for the magnetic dissipation rates ad hoc. In this section, we will fit the prediction for $\chi(\omega)$ from Eq.~\eqref{Eqss} to experimentally measured susceptibilities at varying temperatures. The good agreement between experimental data and the theoretical fit is in support of our phenomenological theory for hole burning and confirms a linear-in-$T$ growths of the magnetic dissipation, as it is predicted from a phonon bath (c.f. Eq.~\eqref{BEq9}). In addition, the resulting fitting parameters confirm that the resonances are cause by small quantum corrections to the Ising approximation of the order a few microhertz. 

The experimental data was taken from hole burning experiments on a \Li crystal with $x=0.045$. The sample was prepared such that the contact to the environment and thus the phonon linewidth was minimized\cite{Silevitch2017,Schmidt2014}. The measurements were taken at different temperatures increasing from $T=150\text{mK}$ to $T=350$mK in steps of $50$mK. The drive field was constantly held at an amplitude of $h_d=0.3$Oe and frequency $\omega_d=2\pi\times202$Hz. The probe field locked at an amplitude of $h_p=20$mOe and detuned from the drive field by a few microhertz, $\delta\omega=\omega_p-\omega_d\in [-5,5]$mHz. 

In order to keep the fitting procedure as simple as possible and to minimize the number of free parameters, we assume that the Fano resonance is caused by a single $\Lambda$-scheme, which reduces Eq.~\eqref{Eqss} to three states. Without loss of generality we set $l=1, m=3$ and use the fitting function
\eq{EqFit}{
\chi(\delta\omega)&=&-\beta+\alpha\left(\frac{|\mu_{1,3}|^2\left(i\gamma-\nu+\frac{H_d^2|\mu_{2,3}|^2}{4\eta+6i\gamma}\right)}{H_d^2|\mu_{2,3}|^2-(i\gamma-\nu)(6i\gamma-4(\eta+\nu))}\right.\nonumber\\
&&+\left.\frac{|\mu_{2,3}|^2\left(i\gamma+\nu-2\delta\omega+\frac{H_d^2|\mu_{1,3}|^2}{4(\eta+\nu+\delta\omega)+6i\gamma}\right)}{H_d^2|\mu_{1,3}|^2-(i\gamma+\nu-2\delta\omega)(6i\gamma-4(\eta+\delta\omega))}\right).\ \ \ \ \ \ \
}
The parameters $\alpha$ and $\beta$ are added in order to take into account the experimental measurement procedure, in which the asymptotic behavior (at large detunings $\delta\omega$) of the Fano signal is normalized and isolated from a temperature dependent background signal. We model the dissipation rates to increase linearly with temperature $\gamma=\gamma_0 T$ and insert $\nu=\delta\omega-\epsilon$, $\eta=\omega_d-\Delta$. The energies $\epsilon, \Delta$ again correspond to the quasi-degenerate, quantum energy splitting and the Ising level splitting, respectively. The drive field amplitude $H_d=5.7$MHz corresponds to $h_d=0.3$Oe. 
\begin{figure}\center\includegraphics[width=\linewidth]{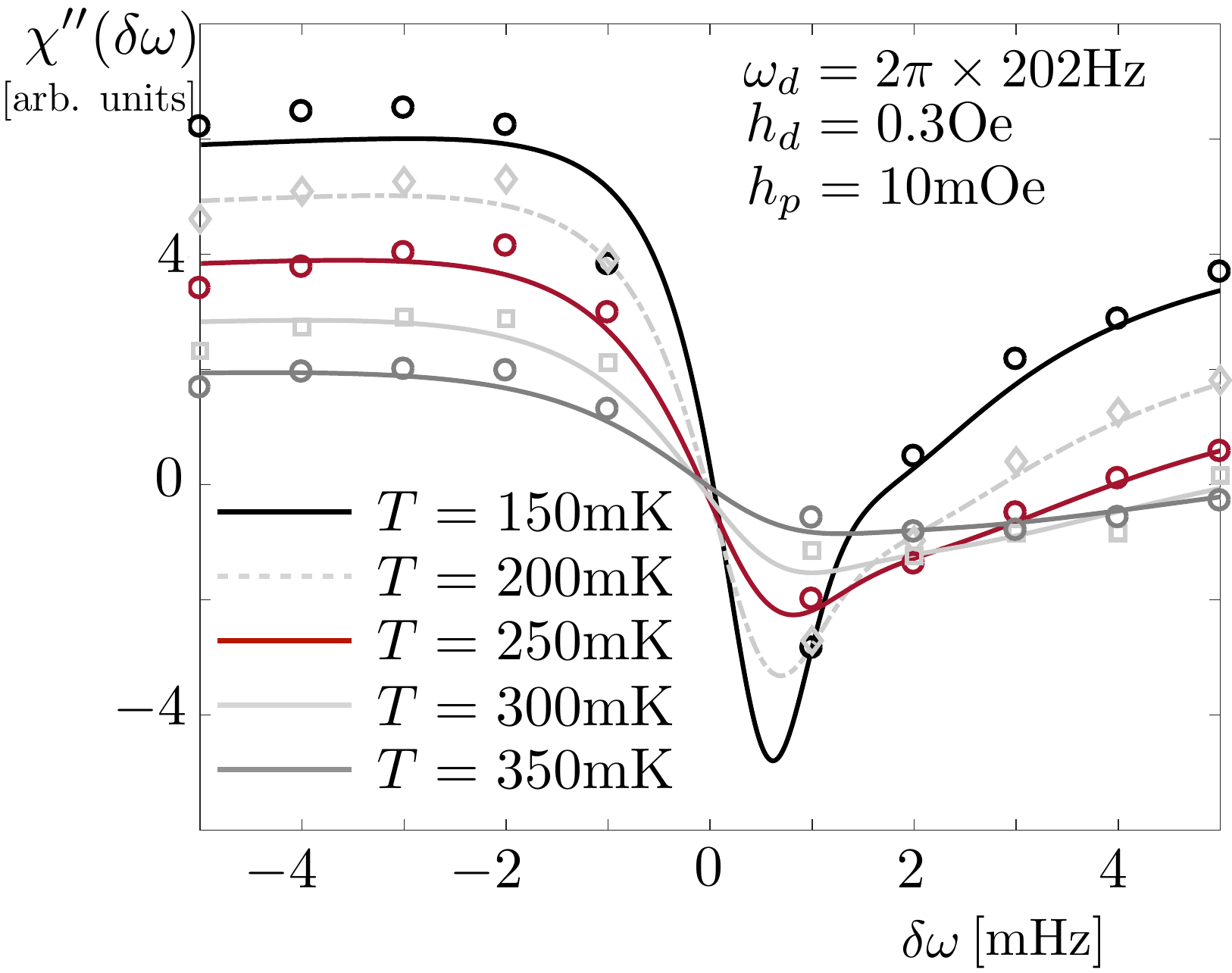}\caption{Comparing experimental data for the imaginary part of the magnetic susceptibility from a \Li sample with $x=0.045$ and theoretical predictions from a single $\Lambda$-scheme in Eq.~\eqref{Eqss} yields very good agreement. The experimental data is represented by markers (circles, diamonds and squares) and was taken for varying probe field detuning $\delta\omega=\omega_p-\omega_d\in2\pi\times [-5, 5]$mHz. The temperature of the sample varies from curve to curve, ranging from $T=150$mK to $T=350$mK. The lines are predictions from Eq.~\eqref{Eqss} for a single $\Lambda$-scheme (without loss of generality $l=1$, $m=3$ with transition matrix elements $\mu_{1,3}=2.3\times10^{-5}$, $\mu_{2,3}=7.1\times10^{-6}$, quasi-degeneracy $\epsilon=E_{12}=22$mHz, Ising detuning $\eta=\omega_d-\Delta=21.4$mHz and $T$-linear dissipation rate $\gamma=0.48$mHz$\times\frac{T}{150\text{mK}}$. The comparison demonstrates, that the experimentally observed signal is very well explained already on the basis of a single $\Lambda$-scheme, and with energy levels and transition matrix elements, which agree well with our predictions for small magnetic clusters in \Li. The linear temperature dependence of the dissipation rate is in agreement with acoustic phonons at very small energy differences $~\sim \omega_d$.}
\label{FigTheExp}\end{figure}

The comparison between the theoretical fit and the experimental data is shown in Fig.~\ref{FigTheExp}. It shows very good agreement between experiment and the prediction from a single $\Lambda$-scheme. All curves share the same transition matrix elements $\mu_{1,3}=2.3\times10^{-5}, \mu_{2,3}=7.1\times10^{-6}$, energy levels $\epsilon=22$mHz, $\omega_p-\Delta=21.4$mHz and a linearly increasing decay rate $\gamma=0.48$mHz$\times\frac{T}{150\text{mK}}$. The parameters $\alpha, \beta$ display a nonlinear temperature dependence and we find $\alpha=(247, 244, 212, 168, 116)$ and $\beta=(5,3.8,2.7,1.8,1.1)$ for the temperatures $T=(150,200,250,300,350)$mK. The monotonic decrease of these values with temperature is likely to be caused by the general decrease in the measured signal for the susceptibility for increasing temperatures.

\section{Anti-hole burning via driven lattice vibrations}
The dissipation experienced by the magnetic moments in the \Li samples is not easy to control experimentally\cite{Silevitch2017, Schmidt2014}. The dissipation rate depends not only on the density of states and the (thermal) occupation of the phonon modes but is also strongly affected by the system-environment coupling, see Eq.~\eqref{BEq8a}. Here we suggest a mechanism to manipulate dissipation, which is experienced by the magnetic degrees of freedom, in a more controllable and purposeful way by energy resolved heating. The basic idea behind this approach is to drive the lattice vibrations, i.e., the phonon modes, in a \Li crystal monochromatically with frequency $\nu_d$. In the low frequency regime $\nu_d\eqsim 0.1-10$kHz, were phonon-phonon scattering is weak, only phonon modes, which are resonant with the drive are heated up. The corresponding nonequilibrium steady state of the lattice is well described by an energy dependent effective temperature $T_{\text{eff}}(E)=T+\Delta T\delta(E-\nu_d)$, which is peaked at the drive frequency but otherwise flat and given by the initial temperature of the sample $T$. 

For the magnetic degrees of freedom, this nonequilibrium state of the lattice translates towards energy dependent dissipation rates $\gamma(E)$, which are as well peaked at $\nu_d$. Magnetic transitions at energy $E=\nu_d$ will therefore experience much stronger dissipation that other transitions at higher or lower energies. In our \Li level scheme, this allows one to target the explicit suppression or elimination of those $\Lambda$-schemes, which display transitions at $\nu_d$. For sufficiently strong phonon driving, the spectral holes at the corresponding frequency will disappear completely. The observation of this ``anti-hole burning'' would be strongly supportive of our theory and yields a further knob to manipulate the low energy physics in \Li samples.

The dependence of the Fano signal on the phonon degrees of freedom has been observed in previous experiments\cite{Silevitch2017}. As we pointed out, reducing the phonon linewidth via decoupling the lattice from the environment is crucial for observing Fano resonances. The coupling to the environment, however, is not an easily tunable parameter. Similarly, the dependence of the magnetic susceptibility on the temperature of the sample, which is a measure of the total phonon occupation, has been studied and a strong reduction of the Fano resonances has been observed for increasing temperature (see Fig.~\ref{FigTheExp}). Temperature, however, increases the dissipation rate uniformly without frequency resolution. 

In order to estimate the effect of acoustic driving on the lattice degrees of freedom, we consider a simple toy model for phonon modes subject to external driving, which is given by the Hamiltonian
\eq{DEq1}{
H_{\text{ph}}=\sum_{\vec{k}} c|\vec{k}| b^\dagger_{\vec{k}}b^{\phantom{\dagger}}_{\vec{k}}+ A\cos(\nu_d t)(b^{\phantom{\dagger}}_{\vec{k}}+b^{\dagger}_{\vec{k}}).
}
Assuming linear sound absorption with amplitude $A$, the coherent drive couples linearly to the bosonic phonon creation and annihilation operators $b^{\dagger}_{\vec{k}}, b^{\phantom{\dagger}}_{\vec{k}}$ and for weak driving $A\le \nu_d$ one can apply the rotating wave approximation, which yields 
\eq{DEq2}{
\tilde{H}_{\text{ph}}=\sum_{\vec{k}} \tilde{\omega}_{\vec{k}} b^\dagger_{\vec{k}}b^{\phantom{\dagger}}_{\vec{k}}+ \frac{A}{2}(b^{\phantom{\dagger}}_{\vec{k}}+b^{\dagger}_{\vec{k}}),
}
with $\tilde{\omega}_{\vec{k}}=c|\vec{k}|-\nu_d$ and $A=Fu$. The force of the drive $F=ma$ is the product of acceleration of the atoms by the sound waves $a$ and their mass $m$. Realistic values are between $a=0.5-10 g$\cite{Miglio1993}. Together with the phonon matrix element $u=\frac{1}{\sqrt{2m\omega}}$ and the mass of Ho atoms, one reaches Rabi frequencies of $A=4-80$kHz.  

Relaxation of the lattice degrees of freedom, either via coupling to the environment or via phonon-phonon scattering is typically very weak and we approximate it via a Markovian master equation in Lindblad form, which evolves the density matrix $\rho_{\text{ph}}$ of the phonons according to
\eq{DEq3}{
\partial_t\rho_{\text{ph}}&=&i[\rho_{\text{ph}},\tilde{H}_{\text{ph}}]+\sum_{\vec{k}}\gamma_{\downarrow,\vec{k}}\left(b^{\phantom{\dagger}}_{\vec{k}} \rho_{\text{ph}}b^{\dagger}_{\vec{k}}-\frac{1}{2}\left\{b^{\dagger}_{\vec{k}}b^{\phantom{\dagger}}_{\vec{k}},\rho_{\text{ph}}\right\}\right)\ \ \nonumber\\
&&+\sum_{\vec{k}}\gamma_{\uparrow,\vec{k}}\left(b^{\dagger}_{\vec{k}} \rho_{\text{ph}}b^{\phantom{\dagger}}_{\vec{k}}-\frac{1}{2}\left\{b^{\phantom{\dagger}}_{\vec{k}}b^{\dagger}_{\vec{k}},\rho_{\text{ph}}\right\}\right).
}
The rates $\gamma_{\downarrow,\vec{k}}, \gamma_{\uparrow,\vec{k}}$ describe the incoherent annihilation, generation of a phonon at wave vector $\vec{k}$ and will not be specified here. Their ratio $\frac{\gamma_{\downarrow,\vec{k}}}{\gamma_{\uparrow,\vec{k}}}=\exp\left(\frac{c|\vec{k}|}{T}\right)$, however, fulfills detailed balance. 

\begin{figure}[h]
\center\includegraphics[width=\linewidth]{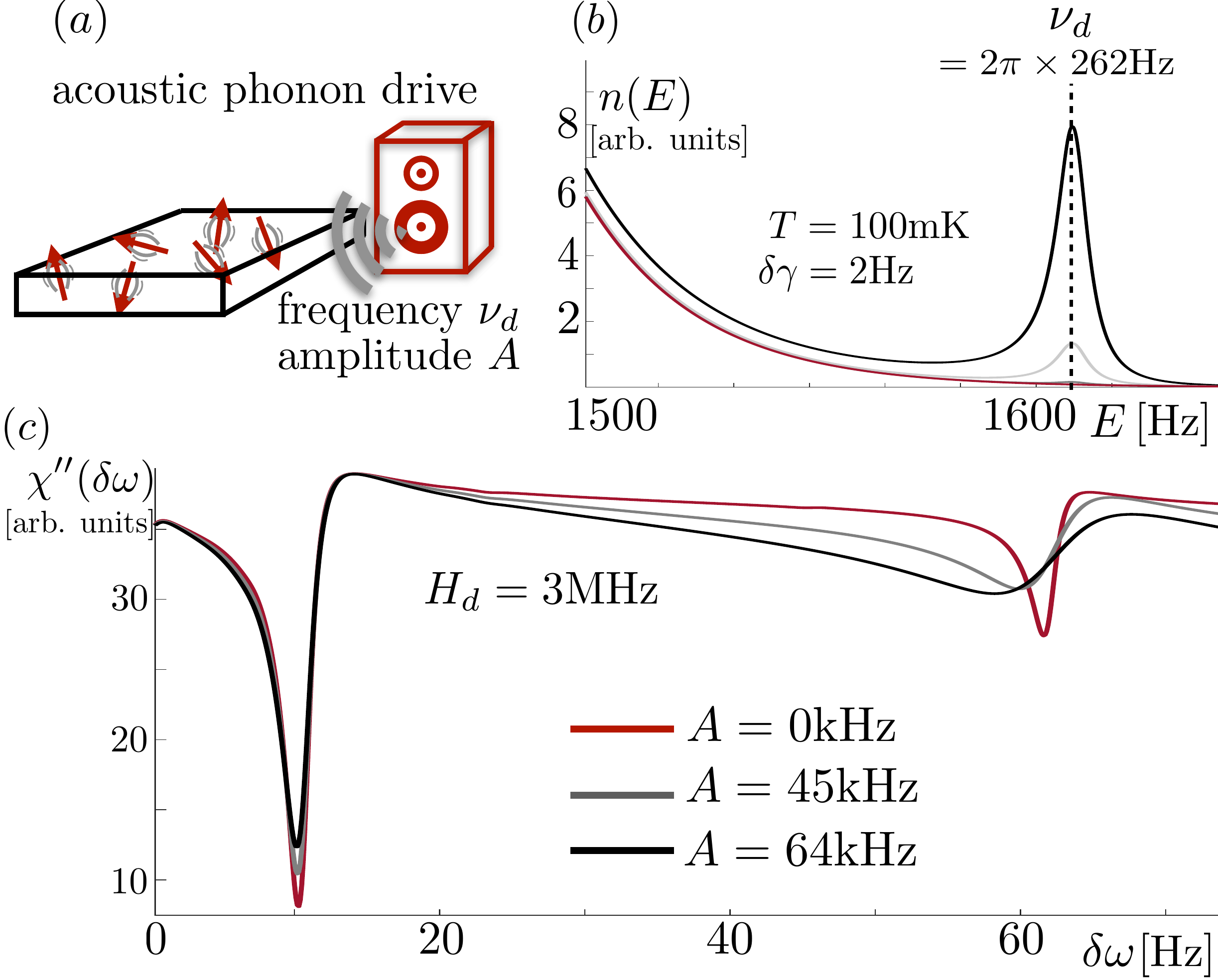}\caption{Manipulating the magnetic dissipation rates in a \Li crystal yields experimental control over the degree to which hole burning (or in general quantum effects) can be observed. Changing the temperature populates or depopulates all low energy lattice degrees of freedom at once. In contrast, monochromatic driving of the lattice, illustrated in (a), can be used to populate only phonon modes in a narrow frequency regime. This is shown in the nonequilibrium phonon distribution $n(E)$ in (b), resulting from a monochromatic drive at frequency $\nu_d=2\pi\times262$ and with variable drive amplitude $A$. The drive populates phonon modes around energy $E\sim \nu_d$, placing a Lorentzian with width $\delta\gamma(\nu_d)$ (inverse phonon lifetime) and height $\frac{A^2}{\delta\gamma(\nu_d)}$ on top of the common Bose-Einstein distribution, cf. Eq.~\eqref{DEq5}. The additional weight in the phonon distribution increases the dissipative magnetic transition rates $\gamma(E)$ at energies $E\sim\nu_d$ close to the drive frequency and leads to dissipation rates described by Eq.~\eqref{DEq6}. $\Lambda$-schemes with energy differences matching $\nu_d$ will thus experience much stronger dissipation and their contribution to hole burning is suppressed. We term this phenomenon anti-hole burning. Its manifestation in the magnetic susceptibility for a \Li sample of $n=12$ magnetic moments is shown in (c). The parameters in (c) are taken from Fig.~\ref{Figfin} (a) and the system is subject to an additional phonon drive at frequency $\nu_d=2\pi\times262$Hz. The phonon drive suppresses the Fano resonance at $\delta\omega\sim 60$Hz but has only little effect on the resonance at $\delta\omega\approx10$Hz. The resonance at $\delta\omega\sim 60$Hz corresponds to a probe frequency $\omega_p=\omega_d+\delta\omega=\nu_d$ and is thus strongly influenced by the phonon drive. The controlled manipulation of magnetic dissipation rates via a monochromatic lattice modulations yields an additional playground for nonequilibrium phenomena in \Li and provides a verification mechanism of the hole burning phenomenology via anti-hole burning.}
\label{Fig7}\end{figure}

Solving the Heisenberg equations of motion $\partial_t n_{\vec{k}}\equiv \partial_t \text{Tr} \left(b^{\dagger}_{\vec{k}}b^{\phantom{\dagger}}_{\vec{k}}\rho_{\text{ph}}\right)$ for the stationary state, $\partial_t n_{\vec{k}}\overset{!}{=}0$ yields
\eq{DEq4}{
n_{\vec{k}}&=&\frac{\gamma_{\uparrow,\vec{k}}}{\gamma_{\downarrow,\vec{k}}-\gamma_{\uparrow,\vec{k}}}+\frac{A^2}{\tilde{\omega}_{\vec{k}}^2+(\gamma_{\downarrow,\vec{k}}-\gamma_{\uparrow,\vec{k}})^2}\\
\Rightarrow n(E)&=& n_B(E)+\frac{A^2}{(E-\nu_d)^2+\delta\gamma(E)^2},\label{DEq5}
}
where we assumed in the second step that the dissipation rates are isotropic and depend only on energy, i.e., $\delta\gamma(E)=\gamma_{\downarrow,\vec{k}}-\gamma_{\uparrow,\vec{k}}$ with $E=c|\vec{k}|$, and we inserted the Bose-Einstein distribution $n_B(E)$. An illustration of the nonequilibriium phonon distribution function in the presence of phonon driving is displayed in Fig.~\ref{Fig7}(b).

%%%

Replacing the Bose distribution in Eq.~\eqref{BEq8a} with the nonequilibrium phonon distribution from Eq.~\eqref{DEq5} and pulling out one factor of $n_B(E)$ from the second part of the equation yields the nonequilibrium magnetic dissipation rate
\eq{DEq6}{
\gamma_{\text{noneq}}(E)=\gamma(E)\left(1+\frac{\nu_d}{T}\frac{A^2}{(E-\nu_d)^2+\delta\gamma(\nu_d)^2}\right).
}
Here we have used the notation $\gamma(E)$ for the equilibrium dissipation rates without phonon driving and approximated $n_B(E)\approx \frac{T}{\nu_d}$ for $T\gg\nu_d$ in the vicinity of the Lorentzian peak. 

Within this simple model, one finds that driving lattice vibrations with a frequency $\nu_d$ and strength $A$ modifies the magnetic dissipation rate by an additional Lorentzian, peaked at $E=\nu_d$ and with maximum $\sim \frac{A^2\nu_d\gamma(\nu_d)}{T\delta\gamma(\nu_d)^2}$ and width $\delta\gamma(\nu_d)$. 

In order to account for the modified dissipation rates in the magnetic susceptibility, one has to replace $\gamma$ in Eq.~\eqref{Eqss} by $\gamma\rightarrow \gamma_{\text{noneq}}(E)$, where $E=E_{lm}$ is the energy of the corresponding transition $|l\rangle\leftrightarrow|m\rangle$. While the complete evaluation of $\chi(\omega)$ becomes complicated with this substitution and can only be performed numerically, we can devise a simple rule of thumb for the modifications due to the phonon drive: since a Fano signal appears only for near resonant transitions $E_{lm}=\omega_{p,d}$, anti-hole burning will be most pronounced at $\omega_p=\nu_d$, i.e. when the phonons are driven close to the probe frequency of the oscillating magnetic field. This behavior is demonstrated via the numerical evaluation of $\chi'(\delta\omega)$ in the presence of phonon driving in Fig.~\ref{Fig7}(c).

Probing anti-hole burning via acoustically driving lattice vibrations should be accessible for most state of the art experiments on \Li and should be able to either confirm or invalidate our present toy model approach. In the case that our predictions survive the experimental reality, the addition of acoustic driving represents a rather simple additional control mechanism for the low energy physics in disordered \Li magnets.

\section{Conclusions}
In this work, we present a numerical analysis of the level structure and magnetic susceptibility of strongly driven \Li samples. It is inspired by experiments that observed spectral hole burning in the susceptibility as the defining characteristic of the antiglass state\cite{Silevitch2017,Reich1987, Ghosh2002a}.

We demonstrate that this spectral hole burning, i.e., Fano resonances in the magnetic susceptibility in \Li can be explained on the basis of small spin clusters ($n=2,3$) and that it can be seen as a consequence of quantum corrections to the common Ising approximation. The Fano resonances persist also when extending the system to the many-body regime, in our numerical study represented by \Li samples of $n=12$ spins and a dilution of $x=0.04$. In the presence of more and more spins, hole burning is caused by interference between many-body quantum spin levels and can be observed at surprisingly low energies and driving frequencies. The crucial requirement for its observation at low frequencies is, however, dissipation rates, i.e., phonon lifetimes, which are of the order of the driving frequencies or even smaller. This requires strong isolation of the system from its environment as also reported in experiments\cite{Silevitch2017}. We also propose an experiment that would test our hypothesis through the excitation of the phonon degrees of freedom crucial for the observation of hole burning. Indeed, we expect that exciting phonons at the appropriate frequency provides an accessible mean of control of hole-burning.

The explanation of the Fano resonances, originating from almost isolated, many-body $\Lambda$-schemes, which are in turn caused by quantum corrections to the classical Ising approximation in \Li without transverse field paves the way further studies on the emergence of quantum effects in strongly diluted and strongly frustrated \Li samples and the role these corrections play for the low energy phase diagram, i.e., for a glass or anti-glass phase. 

Possible further directions include the effect of a transverse field, whose common effect on \Li samples is to introduce or increase quantum effects\cite{Tabei2008,Brooke779,Brooke2001} and to investigate refinements to the model from hyperfine effects in the presence of large transverse fields\cite{Schechter2005}. 

\begin{acknowledgements} 
This work was partially supported by the Department of Energy under Grant No. de-sc0019166. M.~B. acknowledges support from the Alexander von Humboldt foundation. T. F. R. acknowledges support from US Department of Energy Basic Energy Sciences Award DE-SC0014866. We thank Gabriel Aeppli and Markus M\"uller for fruitful discussions.
\end{acknowledgements}

\appendix
\section{Crystal field Hamiltonian}\label{CFH}
The actual form of the crystal field Hamiltonian $H_{\text{cf}}$ for a given electron configuration depends on the symmetries of the crystal (space group $C^6_{4h}$--$I4_1/a$ for \Li)  and the ground state manifold of the ion ($^5I_8$). It is commonly expressed in terms of the so-called Stevens operators \cite{Stevens_1952,Bleaney_1953} $O^\alpha_n$. The crystal field Hamiltonian is
\eq{EqA1}{
H_{\text{cf}}=\sum_{n,\alpha}B_n^\alpha O^\alpha_n
}
and for \Li only $n=0,2,4,6$ have nonzero coefficients $B_n^\alpha$ \cite{GingrasModel}. In terms of the angular momentum operators $J_{\pm}, J_z$ and the total angular momentum $J^2$, the list of relevant Stevens operators is\cite{GingrasModel, Chak2004} 
\eq{EqA2}{
O_2^{0\phantom{C}}&=&3J_z^2-J^2,\\
O_4^{0\phantom{C}}&=&3J^4-6J^2\left(1+5J_z^2\right)+5J_z^2\left(5+7J_z^2\right),\\
O_4^{4C}&=&\frac{1}{2}\left(J_+^4+J_-^4\right),\\
O_4^{4S}&=&\frac{1}{2i}\left(J_+^4-J_-^4\right),\\
O_6^{0\phantom{C}}&=&-5 J^6+5 J^4 \left(21 J_z^2+8\right)-15 J^2 \left(21 J_z^4+35 J_z^2+4\right)\nonumber\\
&&+21 J_z^2 \left(11 J_z^4+35 J_z^2+14\right),\\
O_6^{4C}&=&\frac{1}{4}\left\{\left(J_+^4+J_-^4\right),\left(11J_z^2-J^2-38\right)\right\},\\
O_6^{4S}&=&\frac{1}{4i}\left\{\left(J_+^4-J_-^4\right),\left(11J_z^2-J^2-38\right)\right\}.
}
The numerical values for the parameters $B_n^\alpha$ are taken from inelastic neutron scattering experiments on LiHoF$_4$\cite{Ronnow2007}. The exact eigenstates of $H_{\text{cf}}$ for this data indeed show up to numerical precision a degenerate ground state doublet and a single excited state at $\Delta E_1=10.849$K with the next excited state at $\Delta E_2=32.136$K above the ground state manifold. 
\section{Strongly two-mode driven two-level systems}\label{StrongDr}
In this section we consider a two-mode driven two level system and show that in the limit of strong drive amplitudes the system performs Rabi oscillations with a strongly suppressed, effective Rabi frequency. The two-level system is described by the Hamiltonian
\eq{SA1}{
H(t)=\frac{1}{2}\left(\Omega_1 \cos(\omega t)+\Omega_2\cos((\omega+\delta)t)\right)\sigma_z+\frac{\Delta}{2}\sigma_x,\ \ \ 
}
with a hierarchy of scales $\Omega_1>\Omega_2\gg \omega\gg \Delta,\delta$. The common rotating wave approximation is not applicable since both Rabi frequencies are much larger than any other energy scale. 

We follow the approach outlined in Ref.~\cite{Nori2007,Grifoni2010} and transform the Hamiltonian into a rotating frame $\tilde H(t)=U^\dagger(t)H(t)U(t)-iU^\dagger(t)\partial_t U(t)$ with $U(t)=\exp\left(-i\sigma_z\left[\frac{\Omega_1}{2\omega}\sin(\omega t)+\frac{\Omega_2}{2(\omega+\delta)}\sin((\omega+\delta)t)\right]\right)$. This yields
\eq{SA2}{
\tilde{H}(t)=\Delta\left[\exp\left(\frac{i\Omega_1\sin(\omega t)}{\omega}+\frac{i\Omega_2\sin((\omega+\delta) t)}{\omega+\delta}\right)\sigma^++\text{ H.c.}\right].\ \ \ \ 
}
The Jacobi-Anger expansion $e^{iz\sin\theta}=\sum_{n=-\infty}^\infty J_n(z) e^{in\theta}$, with the Bessel functions of the first kind $J_n(z)$, of this term yields
\eq{SA3}{\tilde{H}(t)&=&\frac{\Delta}{2} \left(\sigma^+ \sum_{n,m}J_{n-m}\left(\frac{\Omega_1}{\omega}\right)J_m\left(\frac{\Omega_2}{\omega+\delta}\right)e^{i(n\omega+m\delta)t}+\text{ H.c. }\right)\nonumber\\
&=&\sum_{n,m}\left(\frac{\Delta_{m,n}}{2}e^{i(n\omega+m\delta)t}\sigma^+ + \text{H.c. }\right).
}

For strong driving, the effective 'Rabi-frequencies' $\Delta_{m,n}$ are much smaller than the original frequency $|\Delta_{m,n}|\ll |\Delta|$ since $|J_{n}(x)|\sim \sqrt{\frac{\pi}{2x}}$. This enables a rotating wave type approximation in the Floquet frame. The two-dimensional Floquet Hamiltonian corresponding to Eq.~\eqref{SA3} is
\eq{SA4}{
H_{m,n,m',n'}=\delta_{m,m'}\delta_{n,n'}(\omega n+\delta m)+\frac{\Delta_{m-m',n-n'}}{2}(\sigma^++(-1)^{n-n'}\sigma^-).
}
Changes in $n$ are strongly suppressed by the large frequency $\omega$ compared to the Rabi-frequencies. We therefore only consider $n=n'=0$. 
Multiplication with the unitary $U_x=\frac{1}{\sqrt{2}}(\sigma^z+\sigma^x)$ yields a long-range hopping model in Floquet space
\eq{SA6}{
H_{m,0,m',0}=\delta_{m,m'}\delta m+\frac{\Delta_{m-m',0}}{2}\sigma^z.
}

Translating this model back to our original spin model, the detuning $\delta\sim $mHz and the Rabi frequency $\Delta_{m-m',0}\sim \frac{\epsilon\omega_d}{2\pi\sqrt{\Omega_d\Omega_p}}\sim 10^{-2}\epsilon$ where $\epsilon\sim $mHz is a quantum level splitting. This yields incredibly slow Rabi oscillations which do not interfere with the susceptibility at the probe frequency. 

\section{Lattice induced dissipation in the Born-Markov approximation}\label{LatDis}
This section provides a short review over the derivation of phonon induced dissipation rates as shown in Eq.~\eqref{BEq8a}, which were obtained from tracing out the phonon bath in the so-called Born-Markov approximation. We consider a general Hamiltonian of the form $H_{\text{tot}}=H_{\text{mag}}+H_D+H_{\text{mag-ph}}$. Here, $H_{\text{mag}}=\sum_{\alpha}E_\alpha |\alpha\rangle\langle\alpha|$ is the Hamiltonian for the magnetic degrees of freedom, e.g., from Eq.~\eqref{BEq1}, expressed in its eigenbasis and $H_D$ is the Debye-phonon Hamiltonian \eqref{BEq8}. The phonon and magnetic degrees of freedom are coupled via $H_{\text{mag-ph}}$ as shown in Eq.~\eqref{BEq9a}. The time evolution of the total density matrix $\rho_{tot}$, which describes the coupled system of magnetic and phonon modes is given by the von Neumann equation
\eq{App1}{
\partial_t \rho_{\text{tot}}(t)=i[\rho_{\text{tot}}(t),H_{\text{tot}}].
}
It is common to switch to a Dirac representation of the density matrix, $\tilde{\rho}_{\text{tot}}(t)\equiv e^{-it(H_D+H_{\text{mag}})}\rho_{\text{tot}}(t)e^{it(H_D+H_{\text{mag}})}$ and the phonon-magnet coupling $\tilde{H}_{\text{mag-ph}}(t)\equiv e^{it(H_D+H_{\text{mag}})}H_{\text{mag-ph}}e^{-it(H_D+H_{\text{mag}})}$. This yields the equation of motion
\eq{App2}{
\partial_t \tilde \rho_{\text{tot}}(t)=i[\tilde\rho_{\text{tot}}(t),\tilde H_{\text{mag-ph}}(t)].
}
It is formally solved by \eq{App3}{\tilde \rho_{\text{tot}}(t)-\rho(0)=\int_0^t i[\tilde\rho_{\text{tot}}(t'),\tilde H_{\text{mag-ph}}(t')]dt',} which we insert into \eqref{App2} and find
\eq{App4}{
\partial_t \tilde \rho_{\text{tot}}(t)&=&i[\rho_{\text{tot}}(0),\tilde H_{\text{mag-ph}}(t)]\nonumber\\&&-\int_0^t \left[[\tilde\rho_{\text{tot}}(t'),\tilde H_{\text{mag-ph}}(t')],H_{\text{mag-ph}}(t)\right]dt'. 
}

The density matrix of the magnetic degrees of freedom is obtained from $\tilde\rho_{\text{tot}}$ by taking the partial trace over the phonon degrees of freedom, i.e., $\tilde{\rho}_{\text{mag}}(t)=\text{Tr}_{\text{ph}}(\tilde{\rho}_{\text{tot}(t)})$. Assuming that the initial density matrix is a direct product of the magnetic and phonon Hilbert spaces and that it commutes with the $H_{D}$ and $H_{\text{mag}}$, one finds the formally exact expression
\eq{App5}{
\partial_t\tilde\rho_{\text{mag}}(t)=-\text{Tr}_{\text{ph}}\left(\int_0^t \left[[\tilde\rho_{\text{tot}}(t'),\tilde H_{\text{mag-ph}}(t')],H_{\text{mag-ph}}(t)\right]dt'\right).\ \ \ \
}

Within the Born-Markov approximation only terms up to second order in the magnetic-phonon coupling $g$ are taken into account and one assumes that the phonon system relaxes towards its equilibrium on time scales much faster than $g^{-1}$, i.e., the phonon system always remains in its thermal equilibrium state. As a consequence of both approximations the density matrix can be written as an instantaneous product, $\tilde\rho_{\text{tot}}(t')\rightarrow \tilde\rho_{\text{mag}}(t)\otimes \rho_{\text{ph}}(0) $. The trace over the phonon states may now be performed in the eigenbasis of $H_D$, which are product states of the form $\prod_{\vec{k}}|n_{\vec{k}}\rangle$, in bosonic Fock space. 

Approximating the integral via sending the integral bounds to $\pm \infty$, which is a good approximation for a rapidly oscillating kernel and going back to the Schr\"odinger picture for the density matrix $\rho_{\text{mag}}$, one finds 
\eq{App6}{
\partial_t\rho_{\text{mag}}&=&i[\rho_{\text{mag}},H_{\text{mag}}]\\
&&+\sum_{\alpha,\beta}\gamma(E_{\alpha\beta})\left[|\alpha\rangle\langle\beta|\rho_{\text{mag}}|\beta\rangle\langle\alpha| -\frac{1}{2}\left\{|\beta\rangle\langle\beta|,\rho_{\text{mag}}\right\}\right],\nonumber
}
where $\gamma(E_{\alpha\beta})$ is given by Eq.~\eqref{BEq8a}.
\bibliography{Fano}

\begin{thebibliography}{48}
\expandafter\ifx\csname natexlab\endcsname\relax\def\natexlab#1{#1}\fi
\expandafter\ifx\csname bibnamefont\endcsname\relax
  \def\bibnamefont#1{#1}\fi
\expandafter\ifx\csname bibfnamefont\endcsname\relax
  \def\bibfnamefont#1{#1}\fi
\expandafter\ifx\csname citenamefont\endcsname\relax
  \def\citenamefont#1{#1}\fi
\expandafter\ifx\csname url\endcsname\relax
  \def\url#1{\texttt{#1}}\fi
\expandafter\ifx\csname urlprefix\endcsname\relax\def\urlprefix{URL }\fi
\providecommand{\bibinfo}[2]{#2}
\providecommand{\eprint}[2][]{\url{#2}}

\bibitem[{\citenamefont{Chakraborty et~al.}(2004)\citenamefont{Chakraborty,
  Henelius, Kj\o{}nsberg, Sandvik, and Girvin}}]{Chak2004}
\bibinfo{author}{\bibfnamefont{P.~B.} \bibnamefont{Chakraborty}},
  \bibinfo{author}{\bibfnamefont{P.}~\bibnamefont{Henelius}},
  \bibinfo{author}{\bibfnamefont{H.}~\bibnamefont{Kj\o{}nsberg}},
  \bibinfo{author}{\bibfnamefont{A.~W.} \bibnamefont{Sandvik}},
  \bibnamefont{and} \bibinfo{author}{\bibfnamefont{S.~M.}
  \bibnamefont{Girvin}}, \bibinfo{journal}{Phys. Rev. B}
  \textbf{\bibinfo{volume}{70}}, \bibinfo{pages}{144411}
  (\bibinfo{year}{2004}),
  \urlprefix\url{https://link.aps.org/doi/10.1103/PhysRevB.70.144411}.

\bibitem[{\citenamefont{Biltmo and Henelius}(2007)}]{Biltmo2007}
\bibinfo{author}{\bibfnamefont{A.}~\bibnamefont{Biltmo}} \bibnamefont{and}
  \bibinfo{author}{\bibfnamefont{P.}~\bibnamefont{Henelius}},
  \bibinfo{journal}{Phys. Rev. B} \textbf{\bibinfo{volume}{76}},
  \bibinfo{pages}{054423} (\bibinfo{year}{2007}),
  \urlprefix\url{https://link.aps.org/doi/10.1103/PhysRevB.76.054423}.

\bibitem[{\citenamefont{Schechter and Stamp}(2008)}]{Schechter2008}
\bibinfo{author}{\bibfnamefont{M.}~\bibnamefont{Schechter}} \bibnamefont{and}
  \bibinfo{author}{\bibfnamefont{P.~C.~E.} \bibnamefont{Stamp}},
  \bibinfo{journal}{Phys. Rev. B} \textbf{\bibinfo{volume}{78}},
  \bibinfo{pages}{054438} (\bibinfo{year}{2008}),
  \urlprefix\url{https://link.aps.org/doi/10.1103/PhysRevB.78.054438}.

\bibitem[{\citenamefont{Brooke et~al.}(1999)\citenamefont{Brooke, Bitko,
  Rosenbaum, and Aeppli}}]{Brooke779}
\bibinfo{author}{\bibfnamefont{J.}~\bibnamefont{Brooke}},
  \bibinfo{author}{\bibfnamefont{D.}~\bibnamefont{Bitko}},
  \bibinfo{author}{\bibfnamefont{T.~F.} \bibnamefont{Rosenbaum}},
  \bibnamefont{and} \bibinfo{author}{\bibfnamefont{G.}~\bibnamefont{Aeppli}},
  \bibinfo{journal}{Science} \textbf{\bibinfo{volume}{284}},
  \bibinfo{pages}{779} (\bibinfo{year}{1999}), ISSN \bibinfo{issn}{0036-8075},
  \urlprefix\url{https://science.sciencemag.org/content/284/5415/779}.

\bibitem[{\citenamefont{Gingras and Henelius}(2011)}]{Gingras_2011}
\bibinfo{author}{\bibfnamefont{M.~J.~P.} \bibnamefont{Gingras}}
  \bibnamefont{and} \bibinfo{author}{\bibfnamefont{P.}~\bibnamefont{Henelius}},
  \bibinfo{journal}{Journal of Physics: Conference Series}
  \textbf{\bibinfo{volume}{320}}, \bibinfo{pages}{012001}
  (\bibinfo{year}{2011}),
  \urlprefix\url{https://doi.org/10.1088%2F1742-6596%2F320%2F1%2F012001}.

\bibitem[{\citenamefont{{Silevitch} et~al.}(2007)\citenamefont{{Silevitch},
  {Bitko}, {Brooke}, {Ghosh}, {Aeppli}, and {Rosenbaum}}}]{Silevitch2007}
\bibinfo{author}{\bibfnamefont{D.~M.} \bibnamefont{{Silevitch}}},
  \bibinfo{author}{\bibfnamefont{D.}~\bibnamefont{{Bitko}}},
  \bibinfo{author}{\bibfnamefont{J.}~\bibnamefont{{Brooke}}},
  \bibinfo{author}{\bibfnamefont{S.}~\bibnamefont{{Ghosh}}},
  \bibinfo{author}{\bibfnamefont{G.}~\bibnamefont{{Aeppli}}}, \bibnamefont{and}
  \bibinfo{author}{\bibfnamefont{T.~F.} \bibnamefont{{Rosenbaum}}},
  \bibinfo{journal}{\nat} \textbf{\bibinfo{volume}{448}}, \bibinfo{pages}{567}
  (\bibinfo{year}{2007}).

\bibitem[{\citenamefont{Schechter}(2008)}]{Schechter2008a}
\bibinfo{author}{\bibfnamefont{M.}~\bibnamefont{Schechter}},
  \bibinfo{journal}{Physical Review B} \textbf{\bibinfo{volume}{77}},
  \bibinfo{pages}{020401} (\bibinfo{year}{2008}).

\bibitem[{\citenamefont{Tabei et~al.}(2008{\natexlab{a}})\citenamefont{Tabei,
  Vernay, and Gingras}}]{Tabei2008a}
\bibinfo{author}{\bibfnamefont{S.~M.~A.} \bibnamefont{Tabei}},
  \bibinfo{author}{\bibfnamefont{F.}~\bibnamefont{Vernay}}, \bibnamefont{and}
  \bibinfo{author}{\bibfnamefont{M.~J.~P.} \bibnamefont{Gingras}},
  \bibinfo{journal}{Physical Review B} \textbf{\bibinfo{volume}{77}},
  \bibinfo{pages}{014432} (\bibinfo{year}{2008}{\natexlab{a}}).

\bibitem[{\citenamefont{{Ghosh} et~al.}(2003)\citenamefont{{Ghosh},
  {Rosenbaum}, {Aeppli}, and {Coppersmith}}}]{Ghosh2003}
\bibinfo{author}{\bibfnamefont{S.}~\bibnamefont{{Ghosh}}},
  \bibinfo{author}{\bibfnamefont{T.~F.} \bibnamefont{{Rosenbaum}}},
  \bibinfo{author}{\bibfnamefont{G.}~\bibnamefont{{Aeppli}}}, \bibnamefont{and}
  \bibinfo{author}{\bibfnamefont{S.~N.} \bibnamefont{{Coppersmith}}},
  \bibinfo{journal}{Nature} \textbf{\bibinfo{volume}{425}}, \bibinfo{pages}{48}
  (\bibinfo{year}{2003}).

\bibitem[{\citenamefont{Bitko et~al.}(1996)\citenamefont{Bitko, Rosenbaum, and
  Aeppli}}]{Bitko1996}
\bibinfo{author}{\bibfnamefont{D.}~\bibnamefont{Bitko}},
  \bibinfo{author}{\bibfnamefont{T.~F.} \bibnamefont{Rosenbaum}},
  \bibnamefont{and} \bibinfo{author}{\bibfnamefont{G.}~\bibnamefont{Aeppli}},
  \bibinfo{journal}{Physical Review Letters} \textbf{\bibinfo{volume}{77}},
  \bibinfo{pages}{940} (\bibinfo{year}{1996}).

\bibitem[{\citenamefont{Reich et~al.}(1990)\citenamefont{Reich, Ellman, Yang,
  Rosenbaum, Aeppli, and Belanger}}]{Reich1990}
\bibinfo{author}{\bibfnamefont{D.~H.} \bibnamefont{Reich}},
  \bibinfo{author}{\bibfnamefont{B.}~\bibnamefont{Ellman}},
  \bibinfo{author}{\bibfnamefont{J.}~\bibnamefont{Yang}},
  \bibinfo{author}{\bibfnamefont{T.~F.} \bibnamefont{Rosenbaum}},
  \bibinfo{author}{\bibfnamefont{G.}~\bibnamefont{Aeppli}}, \bibnamefont{and}
  \bibinfo{author}{\bibfnamefont{D.~P.} \bibnamefont{Belanger}},
  \bibinfo{journal}{Phys. Rev. B} \textbf{\bibinfo{volume}{42}},
  \bibinfo{pages}{4631} (\bibinfo{year}{1990}),
  \urlprefix\url{https://link.aps.org/doi/10.1103/PhysRevB.42.4631}.

\bibitem[{\citenamefont{{Ghosh} et~al.}(2002)\citenamefont{{Ghosh},
  {Parthasarathy}, {Rosenbaum}, and {Aeppli}}}]{Ghosh2002a}
\bibinfo{author}{\bibfnamefont{S.}~\bibnamefont{{Ghosh}}},
  \bibinfo{author}{\bibfnamefont{R.}~\bibnamefont{{Parthasarathy}}},
  \bibinfo{author}{\bibfnamefont{T.~F.} \bibnamefont{{Rosenbaum}}},
  \bibnamefont{and} \bibinfo{author}{\bibfnamefont{G.}~\bibnamefont{{Aeppli}}},
  \bibinfo{journal}{Science} \textbf{\bibinfo{volume}{296}},
  \bibinfo{pages}{2195} (\bibinfo{year}{2002}).

\bibitem[{\citenamefont{Schmidt et~al.}(2014)\citenamefont{Schmidt, Silevitch,
  Aeppli, and Rosenbaum}}]{Schmidt2014}
\bibinfo{author}{\bibfnamefont{M.~A.} \bibnamefont{Schmidt}},
  \bibinfo{author}{\bibfnamefont{D.~M.} \bibnamefont{Silevitch}},
  \bibinfo{author}{\bibfnamefont{G.}~\bibnamefont{Aeppli}}, \bibnamefont{and}
  \bibinfo{author}{\bibfnamefont{T.~F.} \bibnamefont{Rosenbaum}},
  \bibinfo{journal}{Proceedings of the National Academy of Sciences}
  \textbf{\bibinfo{volume}{111}}, \bibinfo{pages}{3689} (\bibinfo{year}{2014}).

\bibitem[{\citenamefont{Cooke et~al.}(1975)\citenamefont{Cooke, Jones, Silva,
  and Wells}}]{Cooke_1975}
\bibinfo{author}{\bibfnamefont{A.~H.} \bibnamefont{Cooke}},
  \bibinfo{author}{\bibfnamefont{D.~A.} \bibnamefont{Jones}},
  \bibinfo{author}{\bibfnamefont{J.~F.~A.} \bibnamefont{Silva}},
  \bibnamefont{and} \bibinfo{author}{\bibfnamefont{M.~R.} \bibnamefont{Wells}},
  \bibinfo{journal}{Journal of Physics C: Solid State Physics}
  \textbf{\bibinfo{volume}{8}}, \bibinfo{pages}{4083} (\bibinfo{year}{1975}),
  \urlprefix\url{https://doi.org/10.1088%2F0022-3719%2F8%2F23%2F021}.

\bibitem[{\citenamefont{Mennenga et~al.}(1984)\citenamefont{Mennenga, de~Jongh,
  Huiskamp, and Laursen}}]{MENNENGA198448}
\bibinfo{author}{\bibfnamefont{G.}~\bibnamefont{Mennenga}},
  \bibinfo{author}{\bibfnamefont{L.}~\bibnamefont{de~Jongh}},
  \bibinfo{author}{\bibfnamefont{W.}~\bibnamefont{Huiskamp}}, \bibnamefont{and}
  \bibinfo{author}{\bibfnamefont{I.}~\bibnamefont{Laursen}},
  \bibinfo{journal}{Journal of Magnetism and Magnetic Materials}
  \textbf{\bibinfo{volume}{44}}, \bibinfo{pages}{48 } (\bibinfo{year}{1984}),
  ISSN \bibinfo{issn}{0304-8853}.

\bibitem[{\citenamefont{Hansen et~al.}(1975)\citenamefont{Hansen, Johansson,
  and Nevald}}]{Hansen1975}
\bibinfo{author}{\bibfnamefont{P.~E.} \bibnamefont{Hansen}},
  \bibinfo{author}{\bibfnamefont{T.}~\bibnamefont{Johansson}},
  \bibnamefont{and} \bibinfo{author}{\bibfnamefont{R.}~\bibnamefont{Nevald}},
  \bibinfo{journal}{Phys. Rev. B} \textbf{\bibinfo{volume}{12}},
  \bibinfo{pages}{5315} (\bibinfo{year}{1975}),
  \urlprefix\url{https://link.aps.org/doi/10.1103/PhysRevB.12.5315}.

\bibitem[{\citenamefont{{Brooke} et~al.}(2001)\citenamefont{{Brooke},
  {Rosenbaum}, and {Aeppli}}}]{Brooke2001}
\bibinfo{author}{\bibfnamefont{J.}~\bibnamefont{{Brooke}}},
  \bibinfo{author}{\bibfnamefont{T.~F.} \bibnamefont{{Rosenbaum}}},
  \bibnamefont{and} \bibinfo{author}{\bibfnamefont{G.}~\bibnamefont{{Aeppli}}},
  \bibinfo{journal}{\nat} \textbf{\bibinfo{volume}{413}}, \bibinfo{pages}{610}
  (\bibinfo{year}{2001}).

\bibitem[{\citenamefont{Quilliam et~al.}(2008)\citenamefont{Quilliam, Meng,
  Mugford, and Kycia}}]{QuilliamSG}
\bibinfo{author}{\bibfnamefont{J.~A.} \bibnamefont{Quilliam}},
  \bibinfo{author}{\bibfnamefont{S.}~\bibnamefont{Meng}},
  \bibinfo{author}{\bibfnamefont{C.~G.~A.} \bibnamefont{Mugford}},
  \bibnamefont{and} \bibinfo{author}{\bibfnamefont{J.~B.} \bibnamefont{Kycia}},
  \bibinfo{journal}{Phys. Rev. Lett.} \textbf{\bibinfo{volume}{101}},
  \bibinfo{pages}{187204} (\bibinfo{year}{2008}),
  \urlprefix\url{https://link.aps.org/doi/10.1103/PhysRevLett.101.187204}.

\bibitem[{\citenamefont{{Biltmo} and {Henelius}}(2012)}]{HeneliusNature}
\bibinfo{author}{\bibfnamefont{A.}~\bibnamefont{{Biltmo}}} \bibnamefont{and}
  \bibinfo{author}{\bibfnamefont{P.}~\bibnamefont{{Henelius}}},
  \bibinfo{journal}{Nature Communications} \textbf{\bibinfo{volume}{3}},
  \bibinfo{eid}{857} (\bibinfo{year}{2012}).

\bibitem[{\citenamefont{J\"onsson et~al.}(2007)\citenamefont{J\"onsson,
  Mathieu, Wernsdorfer, Tkachuk, and Barbara}}]{jonsson2007}
\bibinfo{author}{\bibfnamefont{P.~E.} \bibnamefont{J\"onsson}},
  \bibinfo{author}{\bibfnamefont{R.}~\bibnamefont{Mathieu}},
  \bibinfo{author}{\bibfnamefont{W.}~\bibnamefont{Wernsdorfer}},
  \bibinfo{author}{\bibfnamefont{A.~M.} \bibnamefont{Tkachuk}},
  \bibnamefont{and} \bibinfo{author}{\bibfnamefont{B.}~\bibnamefont{Barbara}},
  \bibinfo{journal}{Phys. Rev. Lett.} \textbf{\bibinfo{volume}{98}},
  \bibinfo{pages}{256403} (\bibinfo{year}{2007}),
  \urlprefix\url{https://link.aps.org/doi/10.1103/PhysRevLett.98.256403}.

\bibitem[{\citenamefont{Battison et~al.}(1975)\citenamefont{Battison, Kasten,
  Leask, Lowry, and Wanklyn}}]{Battison_1975}
\bibinfo{author}{\bibfnamefont{J.~E.} \bibnamefont{Battison}},
  \bibinfo{author}{\bibfnamefont{A.}~\bibnamefont{Kasten}},
  \bibinfo{author}{\bibfnamefont{M.~J.~M.} \bibnamefont{Leask}},
  \bibinfo{author}{\bibfnamefont{J.~B.} \bibnamefont{Lowry}}, \bibnamefont{and}
  \bibinfo{author}{\bibfnamefont{B.~M.} \bibnamefont{Wanklyn}},
  \textbf{\bibinfo{volume}{8}}, \bibinfo{pages}{4089} (\bibinfo{year}{1975}),
  \urlprefix\url{https://doi.org/10.1088%2F0022-3719%2F8%2F23%2F022}.

\bibitem[{\citenamefont{Beauvillain et~al.}(1978)\citenamefont{Beauvillain,
  Renard, Laursen, and Walker}}]{Beau1978}
\bibinfo{author}{\bibfnamefont{P.}~\bibnamefont{Beauvillain}},
  \bibinfo{author}{\bibfnamefont{J.~P.} \bibnamefont{Renard}},
  \bibinfo{author}{\bibfnamefont{I.}~\bibnamefont{Laursen}}, \bibnamefont{and}
  \bibinfo{author}{\bibfnamefont{P.~J.} \bibnamefont{Walker}},
  \bibinfo{journal}{Phys. Rev. B} \textbf{\bibinfo{volume}{18}},
  \bibinfo{pages}{3360} (\bibinfo{year}{1978}),
  \urlprefix\url{https://link.aps.org/doi/10.1103/PhysRevB.18.3360}.

\bibitem[{\citenamefont{Tabei et~al.}(2008{\natexlab{b}})\citenamefont{Tabei,
  Gingras, Kao, and Yavors'kii}}]{Tabei2008}
\bibinfo{author}{\bibfnamefont{S.~M.~A.} \bibnamefont{Tabei}},
  \bibinfo{author}{\bibfnamefont{M.~J.~P.} \bibnamefont{Gingras}},
  \bibinfo{author}{\bibfnamefont{Y.-J.} \bibnamefont{Kao}}, \bibnamefont{and}
  \bibinfo{author}{\bibfnamefont{T.}~\bibnamefont{Yavors'kii}},
  \bibinfo{journal}{Phys. Rev. B} \textbf{\bibinfo{volume}{78}},
  \bibinfo{pages}{184408} (\bibinfo{year}{2008}{\natexlab{b}}),
  \urlprefix\url{https://link.aps.org/doi/10.1103/PhysRevB.78.184408}.

\bibitem[{\citenamefont{Tam and Gingras}(2009)}]{Tam2009}
\bibinfo{author}{\bibfnamefont{K.-M.} \bibnamefont{Tam}} \bibnamefont{and}
  \bibinfo{author}{\bibfnamefont{M.~J.~P.} \bibnamefont{Gingras}},
  \bibinfo{journal}{Phys. Rev. Lett.} \textbf{\bibinfo{volume}{103}},
  \bibinfo{pages}{087202} (\bibinfo{year}{2009}),
  \urlprefix\url{https://link.aps.org/doi/10.1103/PhysRevLett.103.087202}.

\bibitem[{\citenamefont{Pollack and Schechter}(2014)}]{Pollack2014}
\bibinfo{author}{\bibfnamefont{Y.~G.} \bibnamefont{Pollack}} \bibnamefont{and}
  \bibinfo{author}{\bibfnamefont{M.}~\bibnamefont{Schechter}},
  \bibinfo{journal}{Phys. Rev. B} \textbf{\bibinfo{volume}{89}},
  \bibinfo{pages}{064414} (\bibinfo{year}{2014}),
  \urlprefix\url{https://link.aps.org/doi/10.1103/PhysRevB.89.064414}.

\bibitem[{\citenamefont{Ancona-Torres et~al.}(2008)\citenamefont{Ancona-Torres,
  Silevitch, Aeppli, and Rosenbaum}}]{Ancona2008}
\bibinfo{author}{\bibfnamefont{C.}~\bibnamefont{Ancona-Torres}},
  \bibinfo{author}{\bibfnamefont{D.~M.} \bibnamefont{Silevitch}},
  \bibinfo{author}{\bibfnamefont{G.}~\bibnamefont{Aeppli}}, \bibnamefont{and}
  \bibinfo{author}{\bibfnamefont{T.~F.} \bibnamefont{Rosenbaum}},
  \bibinfo{journal}{Phys. Rev. Lett.} \textbf{\bibinfo{volume}{101}},
  \bibinfo{pages}{057201} (\bibinfo{year}{2008}),
  \urlprefix\url{https://link.aps.org/doi/10.1103/PhysRevLett.101.057201}.

\bibitem[{\citenamefont{Romitti et~al.}(2019)\citenamefont{Romitti, Zimmer,
  Morais, and Magalhaes}}]{Romitti2019}
\bibinfo{author}{\bibfnamefont{M.~V.} \bibnamefont{Romitti}},
  \bibinfo{author}{\bibfnamefont{F.~M.} \bibnamefont{Zimmer}},
  \bibinfo{author}{\bibfnamefont{C.~V.} \bibnamefont{Morais}},
  \bibnamefont{and} \bibinfo{author}{\bibfnamefont{S.~G.}
  \bibnamefont{Magalhaes}}, \bibinfo{journal}{Phys. Rev. B}
  \textbf{\bibinfo{volume}{99}}, \bibinfo{pages}{014203}
  (\bibinfo{year}{2019}),
  \urlprefix\url{https://link.aps.org/doi/10.1103/PhysRevB.99.014203}.

\bibitem[{\citenamefont{Biltmo and Henelius}(2008)}]{Biltmo2008}
\bibinfo{author}{\bibfnamefont{A.}~\bibnamefont{Biltmo}} \bibnamefont{and}
  \bibinfo{author}{\bibfnamefont{P.}~\bibnamefont{Henelius}},
  \bibinfo{journal}{Phys. Rev. B} \textbf{\bibinfo{volume}{78}},
  \bibinfo{pages}{054437} (\bibinfo{year}{2008}),
  \urlprefix\url{https://link.aps.org/doi/10.1103/PhysRevB.78.054437}.

\bibitem[{\citenamefont{{J{\"o}nsson} et~al.}(2008)\citenamefont{{J{\"o}nsson},
  {Mathieu}, {Wernsdorfer}, {Tkachuk}, and {Barbara}}}]{Jon2008}
\bibinfo{author}{\bibfnamefont{P.~E.} \bibnamefont{{J{\"o}nsson}}},
  \bibinfo{author}{\bibfnamefont{R.}~\bibnamefont{{Mathieu}}},
  \bibinfo{author}{\bibfnamefont{W.}~\bibnamefont{{Wernsdorfer}}},
  \bibinfo{author}{\bibfnamefont{A.~M.} \bibnamefont{{Tkachuk}}},
  \bibnamefont{and}
  \bibinfo{author}{\bibfnamefont{B.}~\bibnamefont{{Barbara}}},
  \bibinfo{journal}{arXiv e-prints} \bibinfo{eid}{arXiv:0803.1357}
  (\bibinfo{year}{2008}), \eprint{0803.1357}.

\bibitem[{\citenamefont{Quilliam et~al.}(2007)\citenamefont{Quilliam, Mugford,
  Gomez, Kycia, and Kycia}}]{Quilliam2007}
\bibinfo{author}{\bibfnamefont{J.~A.} \bibnamefont{Quilliam}},
  \bibinfo{author}{\bibfnamefont{C.~G.~A.} \bibnamefont{Mugford}},
  \bibinfo{author}{\bibfnamefont{A.}~\bibnamefont{Gomez}},
  \bibinfo{author}{\bibfnamefont{S.~W.} \bibnamefont{Kycia}}, \bibnamefont{and}
  \bibinfo{author}{\bibfnamefont{J.~B.} \bibnamefont{Kycia}},
  \bibinfo{journal}{Phys. Rev. Lett.} \textbf{\bibinfo{volume}{98}},
  \bibinfo{pages}{037203} (\bibinfo{year}{2007}),
  \urlprefix\url{https://link.aps.org/doi/10.1103/PhysRevLett.98.037203}.

\bibitem[{\citenamefont{Reich et~al.}(1987)\citenamefont{Reich, Rosenbaum, and
  Aeppli}}]{Reich1987}
\bibinfo{author}{\bibfnamefont{D.~H.} \bibnamefont{Reich}},
  \bibinfo{author}{\bibfnamefont{T.~F.} \bibnamefont{Rosenbaum}},
  \bibnamefont{and} \bibinfo{author}{\bibfnamefont{G.}~\bibnamefont{Aeppli}},
  \bibinfo{journal}{Phys. Rev. Lett.} \textbf{\bibinfo{volume}{59}},
  \bibinfo{pages}{1969} (\bibinfo{year}{1987}),
  \urlprefix\url{https://link.aps.org/doi/10.1103/PhysRevLett.59.1969}.

\bibitem[{\citenamefont{{Silevitch} et~al.}(2019)\citenamefont{{Silevitch},
  {Tang}, {Aeppli}, and {Rosenbaum}}}]{Silevitch2017}
\bibinfo{author}{\bibfnamefont{D.~M.} \bibnamefont{{Silevitch}}},
  \bibinfo{author}{\bibfnamefont{C.}~\bibnamefont{{Tang}}},
  \bibinfo{author}{\bibfnamefont{G.}~\bibnamefont{{Aeppli}}}, \bibnamefont{and}
  \bibinfo{author}{\bibfnamefont{T.~F.} \bibnamefont{{Rosenbaum}}},
  \bibinfo{journal}{Nature Communications} \textbf{\bibinfo{volume}{10}},
  \bibinfo{eid}{4001} (\bibinfo{year}{2019}).

\bibitem[{\citenamefont{Andresen et~al.}(2014)\citenamefont{Andresen,
  Katzgraber, Oganesyan, and Schechter}}]{Andresen}
\bibinfo{author}{\bibfnamefont{J.~C.} \bibnamefont{Andresen}},
  \bibinfo{author}{\bibfnamefont{H.~G.} \bibnamefont{Katzgraber}},
  \bibinfo{author}{\bibfnamefont{V.}~\bibnamefont{Oganesyan}},
  \bibnamefont{and}
  \bibinfo{author}{\bibfnamefont{M.}~\bibnamefont{Schechter}},
  \bibinfo{journal}{Phys. Rev. X} \textbf{\bibinfo{volume}{4}},
  \bibinfo{pages}{041016} (\bibinfo{year}{2014}),
  \urlprefix\url{https://link.aps.org/doi/10.1103/PhysRevX.4.041016}.

\bibitem[{\citenamefont{Schechter and Stamp}(2005)}]{Schechter2005}
\bibinfo{author}{\bibfnamefont{M.}~\bibnamefont{Schechter}} \bibnamefont{and}
  \bibinfo{author}{\bibfnamefont{P.~C.~E.} \bibnamefont{Stamp}},
  \bibinfo{journal}{Phys. Rev. Lett.} \textbf{\bibinfo{volume}{95}},
  \bibinfo{pages}{267208} (\bibinfo{year}{2005}),
  \urlprefix\url{https://link.aps.org/doi/10.1103/PhysRevLett.95.267208}.

\bibitem[{\citenamefont{R\o{}nnow et~al.}(2007)\citenamefont{R\o{}nnow, Jensen,
  Parthasarathy, Aeppli, Rosenbaum, McMorrow, and Kraemer}}]{Ronnow2007}
\bibinfo{author}{\bibfnamefont{H.~M.} \bibnamefont{R\o{}nnow}},
  \bibinfo{author}{\bibfnamefont{J.}~\bibnamefont{Jensen}},
  \bibinfo{author}{\bibfnamefont{R.}~\bibnamefont{Parthasarathy}},
  \bibinfo{author}{\bibfnamefont{G.}~\bibnamefont{Aeppli}},
  \bibinfo{author}{\bibfnamefont{T.~F.} \bibnamefont{Rosenbaum}},
  \bibinfo{author}{\bibfnamefont{D.~F.} \bibnamefont{McMorrow}},
  \bibnamefont{and} \bibinfo{author}{\bibfnamefont{C.}~\bibnamefont{Kraemer}},
  \bibinfo{journal}{Phys. Rev. B} \textbf{\bibinfo{volume}{75}},
  \bibinfo{pages}{054426} (\bibinfo{year}{2007}),
  \urlprefix\url{https://link.aps.org/doi/10.1103/PhysRevB.75.054426}.

\bibitem[{\citenamefont{Sipahigil et~al.}(2012)\citenamefont{Sipahigil,
  Goldman, Togan, Chu, Markham, Twitchen, Zibrov, Kubanek, and
  Lukin}}]{Alp2012}
\bibinfo{author}{\bibfnamefont{A.}~\bibnamefont{Sipahigil}},
  \bibinfo{author}{\bibfnamefont{M.~L.} \bibnamefont{Goldman}},
  \bibinfo{author}{\bibfnamefont{E.}~\bibnamefont{Togan}},
  \bibinfo{author}{\bibfnamefont{Y.}~\bibnamefont{Chu}},
  \bibinfo{author}{\bibfnamefont{M.}~\bibnamefont{Markham}},
  \bibinfo{author}{\bibfnamefont{D.~J.} \bibnamefont{Twitchen}},
  \bibinfo{author}{\bibfnamefont{A.~S.} \bibnamefont{Zibrov}},
  \bibinfo{author}{\bibfnamefont{A.}~\bibnamefont{Kubanek}}, \bibnamefont{and}
  \bibinfo{author}{\bibfnamefont{M.~D.} \bibnamefont{Lukin}},
  \bibinfo{journal}{Phys. Rev. Lett.} \textbf{\bibinfo{volume}{108}},
  \bibinfo{pages}{143601} (\bibinfo{year}{2012}),
  \urlprefix\url{https://link.aps.org/doi/10.1103/PhysRevLett.108.143601}.

\bibitem[{\citenamefont{{Pingault} et~al.}(2017)\citenamefont{{Pingault},
  {Jarausch}, {Hepp}, {Klintberg}, {Becker}, {Markham}, {Becher}, and
  {Atat{\"u}re}}}]{Pingault}
\bibinfo{author}{\bibfnamefont{B.}~\bibnamefont{{Pingault}}},
  \bibinfo{author}{\bibfnamefont{D.-D.} \bibnamefont{{Jarausch}}},
  \bibinfo{author}{\bibfnamefont{C.}~\bibnamefont{{Hepp}}},
  \bibinfo{author}{\bibfnamefont{L.}~\bibnamefont{{Klintberg}}},
  \bibinfo{author}{\bibfnamefont{J.~N.} \bibnamefont{{Becker}}},
  \bibinfo{author}{\bibfnamefont{M.}~\bibnamefont{{Markham}}},
  \bibinfo{author}{\bibfnamefont{C.}~\bibnamefont{{Becher}}}, \bibnamefont{and}
  \bibinfo{author}{\bibfnamefont{M.}~\bibnamefont{{Atat{\"u}re}}},
  \bibinfo{journal}{Nature Communications} \textbf{\bibinfo{volume}{8}},
  \bibinfo{eid}{15579} (\bibinfo{year}{2017}).

\bibitem[{\citenamefont{Fleischhauer et~al.}(2005)\citenamefont{Fleischhauer,
  Imamoglu, and Marangos}}]{Fleischi2005}
\bibinfo{author}{\bibfnamefont{M.}~\bibnamefont{Fleischhauer}},
  \bibinfo{author}{\bibfnamefont{A.}~\bibnamefont{Imamoglu}}, \bibnamefont{and}
  \bibinfo{author}{\bibfnamefont{J.~P.} \bibnamefont{Marangos}},
  \bibinfo{journal}{Rev. Mod. Phys.} \textbf{\bibinfo{volume}{77}},
  \bibinfo{pages}{633} (\bibinfo{year}{2005}),
  \urlprefix\url{https://link.aps.org/doi/10.1103/RevModPhys.77.633}.

\bibitem[{\citenamefont{Limonov et~al.}(2017)\citenamefont{Limonov, Rybin,
  Poddubny, and Kivshar}}]{Limonov2017}
\bibinfo{author}{\bibfnamefont{M.~F.} \bibnamefont{Limonov}},
  \bibinfo{author}{\bibfnamefont{M.~V.} \bibnamefont{Rybin}},
  \bibinfo{author}{\bibfnamefont{A.~N.} \bibnamefont{Poddubny}},
  \bibnamefont{and} \bibinfo{author}{\bibfnamefont{Y.~S.}
  \bibnamefont{Kivshar}}, \bibinfo{journal}{Nature Photonics}
  \textbf{\bibinfo{volume}{11}}, \bibinfo{pages}{543} (\bibinfo{year}{2017}),
  \urlprefix\url{https://doi.org/10.1038/nphoton.2017.142}.

\bibitem[{\citenamefont{Breuer and Petruccione}(2007)}]{Breuer_P}
\bibinfo{author}{\bibfnamefont{H.}~\bibnamefont{Breuer}} \bibnamefont{and}
  \bibinfo{author}{\bibfnamefont{F.}~\bibnamefont{Petruccione}},
  \emph{\bibinfo{title}{The Theory of Open Quantum Systems}}
  (\bibinfo{publisher}{OUP Oxford}, \bibinfo{year}{2007}), ISBN
  \bibinfo{isbn}{9780199213900},
  \urlprefix\url{https://books.google.co.uk/books?id=DkcJPwAACAAJ}.

\bibitem[{\citenamefont{Silevitch et~al.}(2007)\citenamefont{Silevitch,
  Gannarelli, Fisher, Aeppli, and Rosenbaum}}]{Silevitch2007a}
\bibinfo{author}{\bibfnamefont{D.~M.} \bibnamefont{Silevitch}},
  \bibinfo{author}{\bibfnamefont{C.~M.~S.} \bibnamefont{Gannarelli}},
  \bibinfo{author}{\bibfnamefont{A.~J.} \bibnamefont{Fisher}},
  \bibinfo{author}{\bibfnamefont{G.}~\bibnamefont{Aeppli}}, \bibnamefont{and}
  \bibinfo{author}{\bibfnamefont{T.~F.} \bibnamefont{Rosenbaum}},
  \bibinfo{journal}{Phys. Rev. Lett.} \textbf{\bibinfo{volume}{99}},
  \bibinfo{pages}{057203} (\bibinfo{year}{2007}),
  \urlprefix\url{https://link.aps.org/doi/10.1103/PhysRevLett.99.057203}.

\bibitem[{\citenamefont{Ashhab et~al.}(2007)\citenamefont{Ashhab, Johansson,
  Zagoskin, and Nori}}]{Nori2007}
\bibinfo{author}{\bibfnamefont{S.}~\bibnamefont{Ashhab}},
  \bibinfo{author}{\bibfnamefont{J.~R.} \bibnamefont{Johansson}},
  \bibinfo{author}{\bibfnamefont{A.~M.} \bibnamefont{Zagoskin}},
  \bibnamefont{and} \bibinfo{author}{\bibfnamefont{F.}~\bibnamefont{Nori}},
  \bibinfo{journal}{Phys. Rev. A} \textbf{\bibinfo{volume}{75}},
  \bibinfo{pages}{063414} (\bibinfo{year}{2007}),
  \urlprefix\url{https://link.aps.org/doi/10.1103/PhysRevA.75.063414}.

\bibitem[{\citenamefont{Hausinger and Grifoni}(2010)}]{Grifoni2010}
\bibinfo{author}{\bibfnamefont{J.}~\bibnamefont{Hausinger}} \bibnamefont{and}
  \bibinfo{author}{\bibfnamefont{M.}~\bibnamefont{Grifoni}},
  \bibinfo{journal}{Phys. Rev. A} \textbf{\bibinfo{volume}{81}},
  \bibinfo{pages}{022117} (\bibinfo{year}{2010}),
  \urlprefix\url{https://link.aps.org/doi/10.1103/PhysRevA.81.022117}.

\bibitem[{\citenamefont{{Chin} and {Eastham}}(2006)}]{Chin2006}
\bibinfo{author}{\bibfnamefont{A.}~\bibnamefont{{Chin}}} \bibnamefont{and}
  \bibinfo{author}{\bibfnamefont{P.~R.} \bibnamefont{{Eastham}}},
  \bibinfo{journal}{arXiv e-prints}  (\bibinfo{year}{2006}),
  \eprint{cond-mat/0610544}.

\bibitem[{\citenamefont{{Migliori} et~al.}(1993)\citenamefont{{Migliori},
  {Sarrao}, {Visscher}, {Bell}, {Lei}, {Fisk}, and {Leisure}}}]{Miglio1993}
\bibinfo{author}{\bibfnamefont{A.}~\bibnamefont{{Migliori}}},
  \bibinfo{author}{\bibfnamefont{J.~L.} \bibnamefont{{Sarrao}}},
  \bibinfo{author}{\bibfnamefont{W.~M.} \bibnamefont{{Visscher}}},
  \bibinfo{author}{\bibfnamefont{T.~M.} \bibnamefont{{Bell}}},
  \bibinfo{author}{\bibfnamefont{M.}~\bibnamefont{{Lei}}},
  \bibinfo{author}{\bibfnamefont{Z.}~\bibnamefont{{Fisk}}}, \bibnamefont{and}
  \bibinfo{author}{\bibfnamefont{R.~G.} \bibnamefont{{Leisure}}},
  \bibinfo{journal}{Physica B Condensed Matter} \textbf{\bibinfo{volume}{183}},
  \bibinfo{pages}{1} (\bibinfo{year}{1993}).

\bibitem[{\citenamefont{Stevens}(1952)}]{Stevens_1952}
\bibinfo{author}{\bibfnamefont{K.~W.~H.} \bibnamefont{Stevens}},
  \bibinfo{journal}{Proceedings of the Physical Society. Section A}
  \textbf{\bibinfo{volume}{65}}, \bibinfo{pages}{209} (\bibinfo{year}{1952}),
  \urlprefix\url{https://doi.org/10.1088%2F0370-1298%2F65%2F3%2F308}.

\bibitem[{\citenamefont{Bleaney and Stevens}(1953)}]{Bleaney_1953}
\bibinfo{author}{\bibfnamefont{B.}~\bibnamefont{Bleaney}} \bibnamefont{and}
  \bibinfo{author}{\bibfnamefont{K.~W.~H.} \bibnamefont{Stevens}},
  \bibinfo{journal}{Reports on Progress in Physics}
  \textbf{\bibinfo{volume}{16}}, \bibinfo{pages}{108} (\bibinfo{year}{1953}),
  \urlprefix\url{https://doi.org/10.1088%2F0034-4885%2F16%2F1%2F304}.

\bibitem[{\citenamefont{Tabei et~al.}(2008{\natexlab{c}})\citenamefont{Tabei,
  Gingras, Kao, and Yavors'kii}}]{GingrasModel}
\bibinfo{author}{\bibfnamefont{S.~M.~A.} \bibnamefont{Tabei}},
  \bibinfo{author}{\bibfnamefont{M.~J.~P.} \bibnamefont{Gingras}},
  \bibinfo{author}{\bibfnamefont{Y.-J.} \bibnamefont{Kao}}, \bibnamefont{and}
  \bibinfo{author}{\bibfnamefont{T.}~\bibnamefont{Yavors'kii}},
  \bibinfo{journal}{Phys. Rev. B} \textbf{\bibinfo{volume}{78}},
  \bibinfo{pages}{184408} (\bibinfo{year}{2008}{\natexlab{c}}),
  \urlprefix\url{https://link.aps.org/doi/10.1103/PhysRevB.78.184408}.

\end{thebibliography}

\end{document}